\documentclass[preprint]{aastex}

\def\gtorder{\mathrel{\raise.3ex\hbox{$>$}\mkern-14mu
             \lower0.6ex\hbox{$\sim$}}}
\def\ltorder{\mathrel{\raise.3ex\hbox{$<$}\mkern-14mu
             \lower0.6ex\hbox{$\sim$}}}
\def\proptwid{\mathrel{\raise.3ex\hbox{$\propto$}\mkern-14mu
             \lower0.6ex\hbox{$\sim$}}}
\def\arcsec{\ifmmode '' \else $''$\fi}

\def\arcsecpoint{\ifmmode ''\!. \else $''\!.$\fi}

\def\kms{\ifmmode {\rm km\ s}^{-1} \else km s$^{-1}$\fi}
\def\Msun{\ifmmode {\rm M}_{\odot} \else M$_{\odot}$\fi}
\def\Lsun{\ifmmode {\rm L}_{\odot} \else L$_{\odot}$\fi}
\def\Zsun{\ifmmode {\rm Z}_{\odot} \else Z$_{\odot}$\fi}

\def\ergscm2{ergs\,s$^{-1}$\,cm$^{-2}$}
\def\icm3{{\rm cm}^{-3}}
\def\icm2{{\rm cm}^{-2}}
\def\qo{\ifmmode q_{\rm o} \else $q_{\rm o}$\fi}
\def\Ho{\ifmmode H_{\rm o} \else $H_{\rm o}$\fi}
\def\ho{\ifmmode h_{\rm o} \else $h_{\rm o}$\fi}

\def\vFWHM{\ifmmode v_{\mbox{\tiny FWHM}} \else
            $v_{\mbox{\tiny FWHM}}$\fi}
\def\CCF{\ifmmode F_{\it CCF} \else $F_{\it CCF}$\fi}
\def\ACF{\ifmmode F_{\it ACF} \else $F_{\it ACF}$\fi}
\def\Halpha{\ifmmode {\rm H}\alpha \else H$\alpha$\fi}
\def\Hbeta{\ifmmode {\rm H}\beta \else H$\beta$\fi}
\def\Hgamma{\ifmmode {\rm H}\gamma \else H$\gamma$\fi}
\def\Hdelta{\ifmmode {\rm H}\delta \else H$\delta$\fi}
\def\Lya{\ifmmode {\rm Ly}\alpha \else Ly$\alpha$\fi}
\def\Lyb{\ifmmode {\rm Ly}\beta \else Ly$\beta$\fi}
\def\Lyg{\ifmmode {\rm Ly}\beta \else Ly$\gamma$\fi}

\def\cii{C\,{\sc ii}}
\def\ciii{\ifmmode {\rm C}\,{\sc iii} \else C\,{\sc iii}\fi}
\def\civ{\ifmmode {\rm C}\,{\sc iv} \else C\,{\sc iv}\fi}

\def\o5007{[O\,{\sc iii}]\,$\lambda5007$}

\def\ovi{O\,{\sc vi}}

\def\mgi{Mg\,{\sc i}}
\def\mnii{Mn\,{\sc ii}}
\def\Niii{Ni\,{\sc ii}}
\def\mgii{Mg\,{\sc ii}}

\def\siiv{Si\,{\sc iv}}

\def\siII{Si\,{\sc ii}}

\def\fei{Fe\,{\sc i}}
\def\feii{Fe\,{\sc ii}}
\def\feiii{Fe\,{\sc iii}}

\def\o{\o}
\begin{document}
\title{KECK HIRES OBSERVATIONS OF THE QSO FIRST~J104459.6+365605:\\
        EVIDENCE FOR A LARGE SCALE OUTFLOW}

\author{
Martijn de Kool\footnote{RSAA, Mt. Stromlo Observatory, Cotter Road, Weston ACT 2611, Australia; dekool@mso.anu.edu.au},
Nahum Arav\footnote{IGPP,LLNL, L-413, P.O. Box 808, Livermore, CA 94550}, 
Robert H. Becker$^{2,}$\footnote{Physics Department,
University of California, Davis, CA 95616},
Michael D. Gregg$^{2,3}$, \\ Richard L. 
White\footnote{Space Telescope Science Institute, Baltimore, MD 21218}, 
Sally A. Laurent-Muehleisen$^{2,3}$,
Trevor Price$^{2,3}$
and Kirk T. Korista\footnote{Physics Department, Western Michigan University}
}
\begin{abstract}          
This paper presents an analysis of a Keck HIRES spectrum of the QSO
FIRST~J104459.6+365605, covering the rest wavelength range from 2260
to 2900 \AA.  The line of sight towards the QSO contains two clusters
of outflowing clouds that give rise to broad blue shifted absorption
lines. The outflow velocities of the clouds range from --200 to --1200 \kms\
and from --3400 to --5200 \kms, respectively. The width of the
individual absorption lines ranges from 50 to more than 1000 \kms.

The most prominent absorption lines are those of \mgii, \mgi, and
\feii, and \mnii\ is also present. The low ionization absorption lines
occur at the same velocities as the most saturated \mgii\ lines,
showing that the \feii\ , \mgi\ and \mgii\ line forming regions must
be closely associated. Many absorption lines from excited states of
\feii\ are present, allowing a determination of the population of
several low lying energy levels.  The populations of the excited
levels are found to be considerably smaller than expected for LTE, and
imply an electron density in the \feii\ line forming regions of $n_e
\sim 4 \times 10^3$ cm$^{-3}$.

Modelling the ionization state of the absorbing gas with this value of
the electron density as a constraint, we find that the distance
between the \feii\ and \mgi\ line forming region and the continuum
source is $ \sim 7 \times 10^2$ parsec. From the correspondence in
velocity between the \feii\ , \mgi\ and \mgii\ lines we infer that the
\mgii\ lines must be formed at the same distance.  The \mgii\
absorption fulfills the criteria for Broad Absorption Lines defined by
Weymann et al. (1991). Therefore the distance we find between the
\mgii\ line forming region and the continuum source is surprising,
since BALs are generally thought to be formed in outflows at a much
smaller distance from the nucleus.
\end{abstract}
          
\newpage

\section{INTRODUCTION}

Many QSO rest-frame UV spectra exhibit blue shifted absorption lines
that are indicative of an outflow. The properties of these absorption
lines vary over a wide range, from very broad ($>$ 10,000 \kms)
absorption lines, usually observed in high ionization species like
\civ\ and \ovi\ to fairly narrow ($\sim$ 30 \kms) lines from the same
high ionization species, or from low-ionization species like \feii\
and \cii . The absorption lines are often divided in two classes. The
first class contains the Broad Absorption Lines (BALs)(Weymann,
Turnshek \& Christiansen 1985, Turnshek 1988) which are broad
absorption troughs, and can span a large range of outflow velocities
up to several times $10^4$ \kms\ . The second class refers to
Associated Absorption Lines (AALs) (Weymann et al. 1979, Foltz et al.
1986), which have individual absorption components less than a few
hundred \kms\ wide, and span a range of outflow velocities up to
several thousand \kms .  It is not obvious whether this division
reflects a real bimodal distribution of absorption line properties
caused by two different underlying formation mechanisms. There are
many examples of QSOs that exhibit both types of lines simultaneously,
or have properties that are a mixture between the two classes,
eg. high velocity narrow lines or low velocity broad lines.

It is generally thought that BALs are formed relatively close to the
galactic nucleus ($\ltorder 1$ parsec), although the direct
observational evidence for this is not very strong.  The minimum
distance separating the region in which broad absorption lines arise
(the BALR) from the source of the continuum radiation is set by the
observation that the Broad Emission Lines are often partly or
completely absorbed by the outflow (Turnshek 1988).  This places the
BALR outside the Broad Emission Line Region (BELR), the size of which
is now relatively well established with reverberation mapping
techniques at $\sim 0.1$ pc for an average luminosity QSO.  The
maximum distance of the BALR from the central source is more difficult
to constrain. Adopting typical, if uncertain, physical parameters
for the BAL clouds it can easily be shown that unrealistically large
mass and energy loss rates are implied if the lines are formed more
than a few tens of parsecs from the central engine. Further support
for the conjecture that the BALR is not much larger than the BELR
comes from consideration of the acceleration mechanisms, which
generally require that the flow starts at a distance comparable to the
size of the BELR in order to reach the high speeds observed (e.g. Arav
\& Li 1994).  The physical conditions inside the BALR are difficult to
constrain because of the extremely wide and deep absorption lines in
classical BALQSOs. The width complicates the analysis because the
optical depth at a certain wavelength is the convolution of the column
density as a function of velocity and the multiplet structure of the
line, so that the column density as a function of velocity can only be
derived by a deconvolution.  The large depth makes the derived column
densities very sensitive to partial covering of the continuum light
source.

AALs appear to be more diverse, and could arise from gas located in
the AGN environment, the host galaxy, or intervening gas with a
smaller cosmological redshift than the QSO. Some recent studies of
AALs are Hamann, Netzer \& Shields (2000), Petitjean \& Srianand
(1999), Srianand \& Petitjean (2000), Barlow \& Sargent (1997) and
Ganguly et al. (1999).  The analysis of AALs does not suffer as much
from the problems of blending and saturation encounterd with BALs, and
can give quite detailed results on the physical conditions and
geometry of the absorbing gas, as demonstrated in the references
above. From these studies it has become clear that the great majority
of AALs show the effects of partial covering in their observed
multiplet line ratios, (e.g. Hamann et al. 1997, Arav et al. 1999)
which argues against a cosmological origin, and is most easily
explained if the lines are formed in the AGN environment. However, to
our knowledge no firm limits on the distance of the absorbers to the
active nucleus have been derived so far.

Here we investigate the z=0.700 QSO FIRST~J104459.6+365605 (hereafter
FBQS1044+3656).  This object was discovered in the FIRST Bright Quasar
Survey (White et al 2000) which consists of 636 radio-selected quasars
distributed over $\sim 2700$ square degrees.  The absorption lines in
the spectrum of this QSO have several unique features, allowing us to
constrain the density and ionization parameter in the outflow.  The
spectrum contains absorption lines of \mgi\ , \mgii\ and \feii . Some
of the \feii\ lines arise from low-lying excited states. Most of the
\feii\ lines are not very wide nor heavily saturated, behaving roughly
as expected from their relative oscillator strength, making quantitive
estimates of the column density possible.  The derived relative
populations of several low-lying energy levels of \feii\ allow us to
put significant constraints on the physical conditions in the
absorbing clouds.
 
The characteristic velocity and width of the absorption lines in
FBQS1044+3656 are intermediate between BALs and AALs.  The \mgii\
absorption is very broad, and just fulfills the strict criterion
applied to the \civ\ line by Weymann et al. 1991 to decide whether a
QSO is a BALQSO. The lower ionization lines of \feii\ and \mgi\ are
narrower, and more characteristic of AALs. However, we will argue that
the excellent correspondence in velocity between the strongest \mgii\
components and the \mgi\ and \feii\ lines leave little doubt that they
arise from the same outflowing cloud complex.

Our analysis is similar to the study of BALQSO Q0059-2735
by Wampler, Chugai \& Petitjean 1995 (hereafter referred to as WCP).
Q0059-2735 is a classical BALQSO with very deep and wide BAL troughs
in both high and low ionization species.  Its spectrum also contains
several narrow, low ionization absorption line systems, such as \feii,
\Niii, and \mnii, at velocities of a few hundred \kms. Another recent
study with some similarities to ours considers narrow absorption line
systems in the BALQSO APM 08279+5255 (Srianand \& Petitjean 2000),
which show absorption from excited states of \cii\ and \siII\ .

\section{Data Acquisition and Reduction}

On December 26, 1998 we used the High Resolution Echelle Spectrometer
(HIRES, Vogt et al.\ 1994) on the Keck~I 10-m telescope to obtain $4
\times 1800$ second exposures of FBQS1044+3656 covering 3825 -
6280\AA\ using a 1\farcs1 wide slit.  The orders overlap up to
5128\AA, beyond which small gaps occur between orders.  The slit was
rotated to the parallactic angle to minimize losses due to
differential atmospheric refraction.  The observing conditions were
excellent with sub-arcsecond seeing and near-photometric skies.  The
spectra were extracted using routines tailored for HIRES (Barlow
2000).  The resolution of the final spectrum is 4 pixels
pixels FWHM, or 8.8 \kms\ in velocity space.

The redshift of FBQS1044+3656 (z=0.700) was determined by fitting a
Gaussian to the \mgii\ broad emission line. Because the blue wing of
the \mgii\ BEL is strongly affected by absorption, the redshift of the
QSO rest frame is uncertain by several hundred \kms\ . Since all outflow
velocities quoted in this paper are calculated from the difference in
redshift between the absorbing gas and the QSO rest frame, they are
uncertain by the same amount.

\section{Analysis}

In this section we will analyze the absorption lines in the spectrum
of FBQS1044+3656, deriving the column densities of the observed
species.  In the next section, these column densities will be used to
derive the physical condition in the outflow.  An overview of the
HIRES spectrum is shown in figure \ref{totspec}. A low
resolution spectrum of FBQS1044+3656 covering a larger wavelength
range has been published in White et al. 2000.

\subsection{\mgi\ lines}

The \mgi\ line at $\lambda$ 2852.96 \AA\ is a singlet transition from
the ground level. The line profile as a function of velocity is shown
in Figure~\ref{mgplot}, together with the profile of the \mgii\
absorption. The velocity scale of the \mgii\ absorption is based on
the rest wavelength of the red component of the \mgii\ doublet.

We divide the \mgi\ absorption features into two clusters, the first
ranging from $-100$ to $-1400$ \kms\ (cluster 1), the second from
$-3500$ to $-4100$ \kms\ (cluster 2). Some of the components in the
low-velocity cluster are not connected by any absorption, so this
definition of these two clusters is purely ad hoc at this point. It
will turn out however that our division appears to have a physical
basis since the components of cluster 2 exhibit significant absorption
from excited levels of \feii\ , whereas the detection of absorption
from excited levels from the components of cluster 1 is only
marginal. This indicates that the physical conditions differ
between the two clusters. In each cluster we have identified the two
strongest components so that we can refer to these later when
identifying the corresponding absorption lines of other species. Based
on the assumed redshift of 0.700 for the QSO, these components
correspond to an outflow speed of -205, -1227, -3557 and -3860 km
s$^{-1}$ for components A,B,C and D respectively.

A lower limit on the column density of \mgi\ can be obtained in the
optically thin approximation, valid if the line is fully resolved and
the absorbing gas is covering the entire continuum source:
\begin{equation} {N_{ion}} \geq \frac{
{m_ec} }{ \pi\/{e}^2 } ~ \frac{1}{\lambda_{{o}}{f_{lu}} }
\int {\tau\/(v) dv} ,\label{coldeq} 
\end{equation}
The \mgi\ column density derived in this way is $1.4 \times 10^{13}
\icm2$ for cluster 1 and $1.0 \times 10^{13} \icm2$ for cluster 2.
From the analysis of the \feii\ lines below, we will find that the
\feii\ absorbing clouds do not cover the entire continuum source.  It
is not clear if the same applies to the \mgi\ line, since the
continuum at the position of the \feii\ lines contains a large
fraction of broad \feii\ emission which comes from a much more
extended region than the continuum from the accretion disk. Thus it is
possible that the outflow covers the accretion disk source completely
but the \feii\ emission line regions only partially, which would lead
to partial covering being observed for absorption lines in the part of
the spectrum where broad \feii\ line emission contributes
significantly to the background source.  Even if the same partial
covering would apply to the \mgi\ line, however, this would increase
the \mgi\ column density only by a factor $< 2$ since the \mgi\ lines
are significantly shallower than the saturated \feii\ lines.  Larger
deviations could occur if the covering factor for the \mgi\ producing
clouds is very different from that of the \feii\ producing clouds.
This possibility can not be excluded, but the fact that the \mgi\ line
profile can be used as template for both strong and weak \feii\ lines
(as shown below) argues against it.  The presence of \mgi\ in
FBQS1044+3656 stands in important contrast with the absence of \mgi\
absorption in Q0059-2735, which was used by WCP to obtain an upper
limit on the distance of the clouds to the continuum source.

\subsection{\mgii\ lines}

From the comparison of \mgi\ and \mgii\ absorption as a function of
velocity in Figure \ref{mgplot}, it can be seen that the \mgii\
absorption extends to much higher velocities than \mgi.  There is a
strong component centered around $-4115$ \kms\ that does not
appear in \mgi, and further contiguous absorption is visible up to
$-5700$ \kms.  At even higher velocities, some very shallow absorption
systems at -6500, -7310 and -7410 \kms\ are probably due to \mgii\
since both components of the doublet are visible.

The \mgii\ lines are heavily saturated allowing only a lower limit on
the column density.  Over the velocity range covered by cluster 1 and
2, we estimate lower limits on the \mgii\ column density of
$2.6~10^{14} \icm2 $ and $5.8~10^{14} \icm2 $ respectively, but the
real columns are likely to be significantly higher.  The differences
between the \mgii\ and \mgi\ profiles do not imply that the ionization
state has to change with velocity: the column density of \mgi\
associated with the high velocity \mgii\ components would be too small
to be detected if the \mgi / \mgii\ ratio is the same as in the
strongly saturated \mgii\ components.  This interpretation is
supported by the high-velocity absorption not being readily detectable
in the weakest \feii\ absorption lines, but being just visible in the
strongest \feii\ lines.

\subsection{\feii\ lines}
    
\subsubsection{Analysis using the \mgi\ template \label{tempana}}

Most of the physical constraints on the absorbing gas can be extracted
from the numerous absorption lines of \feii\ in the spectrum of
FBQS1044+3656.  Because of the blending of the \feii\ lines, most of
our analysis must use a template for the column density as a function
of velocity, which we derive from the unblended singlet transition of
\mgi, an approach similar to that of Arav et al. 1999a.  The template
method is based on the assumption that the column density of \feii\ in
each energy level $k$ between velocity $v$ and $v+dv$ is proportional
to the column density of \mgi\ in the same velocity interval:
\begin{equation} 
{{d N_{Fe~II}^k}\over{dv}} = c_k {{d N_{Mg~I}}\over{dv}}
\end{equation}
The column density of \mgi\ is derived from the \mgi\ absorption line
profile using
\begin{equation} 
{{d N_{Mg~I}}\over{dv}}=\frac{
{m_ec} }{ \pi\/{e}^2 } ~ \frac{1}{\lambda_{Mg}{f_{Mg}} }
(-\ln I(v)) \label{mgprof}
\end{equation}
where $I(v)$ is the remaining flux in the normalized line profile, and
$f_{Mg}$ the oscillator strength of the \mgi\ resonance line. By using
Equation \ref{mgprof}, we implicitly assume that partial covering
effects are not important for \mgi\ (see above), and that the
absorption lines are fully resolved. Any quantitative analysis must
make the second assumption, although it should be kept in mind that
the thermal line width of Fe and Mg at the temperatures of a few
thousand degrees that are expected for the line forming regions are
only of order 1-2 \kms\ , i.e. well below the resolution of our
observations.  When the constants of proportionality $c_k$ between the
column densities of \mgi\ and \feii $^k$ are specified, we can
calculate the optical depth in every \feii\ line as a function of
velocity, and construct a model \feii\ absorption line spectrum that
can be compared to the observed spectrum of FBQS1044+3656.

From a first inspection of the spectrum it is clear that the
components of cluster 2 show significant absorption in excited \feii\
lines, whereas the components of cluster 1 show up only very weakly in
these lines. Therefore, the two clusters appear to arise from
physically distinct cloud complexes. To keep the number of free
parameters in our fit manageable, we will assume that the level
populations do not vary between different components within each
cluster, but allow for different level populations between the two
clusters.

The \feii\ lines considered in our study are listed in Table 1.  We
included all lines from the lowest energy ($a^6D$) term, and the
strongest lines from the next two terms ($a^4F$ and $a^4D$) with
lowest levels at approximately 0.2 and 1~eV, respectively.  In total,
we consider 43 lines arising from 10 energy levels.  

The fit was obtained by first choosing values for the $c_k$'s for each
cluster and then comparing the predicted and observed \feii\
absorption line spectra. The values of the $c_k$'s were successively
improved by adjusting them to reproduce the observed flux at the
positions in the spectrum where the specific $c_k$ is best
constrained. To be properly constraining, we required that the parts
of the spectrum used to adjust the fit contain a cluster component
that is not blended with another line, and is not very deep so that
partial covering effects play as small a role as possible. Although
many lines are blended, all levels give rise to some absorption
components that occur in a part of the spectrum where blending is not
a problem.

Note that this method of fitting a model to the observed spectrum does
not try to fit individual {\it lines}, but rather {\it level
populations} and thus attempts to fit all lines from a given lower
level with one parameter, the column density of \feii\ in the state
corresponding to this lower level. Because isolated, medium depth
lines arising from the level are the most accurate indicators of this
column density, these are the main drivers for the fit.

The model resulting from the fit obtained in this way is compared with
the observed spectrum in figure \ref{fe2plot}. The thin line
represents the reconstructed spectrum based on the model parameters in  
Table~2. We express the model parameters in terms of relative level
populations and a total column density.
$\zeta_k$ is the level population per unit statistical weight of level $k$
relative to that of the ground level,
\begin{equation} 
\zeta_k \equiv {{N_{Fe~II}^k / g_k}\over{N_{Fe~II}^0 / g_0}}
\end{equation}
$\zeta_k$ is equal to 1 if the levels are in LTE with $\rm kT >>
E_{k}$. The total \feii\ column density  is $\log
N_{Fe~II}=14.7(0.2)$ for cluster 1 and $\log N_{Fe~II}=15.3(0.1)$ for
cluster 2.  The upper limits and error estimates were obtained by
visually comparing model spectra with the observations and estimating
how much deviation from either 0 or the measured value could be
tolerated.  Systematic effects such as errors in oscillator strength
or continuum level are thus not included.

The heavily blended regions of the spectrum are well reproduced using
the $c_k$'s obtained from the isolated lines, proving the consistency
of the result.  The figure shows that for most of the absorption
components the fit is satisfactory, considering that the problem is
heavily overconstrained and that the number of lines that have to be
reproduced is several times larger than the number of energy levels
included in the fit.  The most significant differences between the
observed spectrum and the model that can not be removed by changing
the model parameters occur in the strongest absorption lines from the
ground level.  These strong lines show absorption at high velocity,
extending beyond the highest velocity seen in cluster 2 of \mgi .
This extra absorption is also seen as the very deep component at
$-4115$ \kms\ in \mgii.  The other major deviation between model and
observation is that the strongest \feii\ lines are much deeper in the
model.  It is impossible to obtain a good fit to the five lines
originating from the ground state simultaneously. The most likely
explanation for this discrepancy is partial covering of the continuum
source, which fills in the bottom of the strongest lines. Note that
the underprediction of flux in the deepest lines is the result of our
choice to optimize the fit for medium depth lines (which are the
weakest lines when considering lines from the ground level).

There are several other sources of uncertainty that may also
contribute to discrepancies between the model and the observed
spectrum. One is that there are still some uncertainties in the
oscillator strengths of the UV \feii\ absorption lines studied here.
Table 1 shows a comparison of these oscillator strengths from
different literature sources. In this work we have used the oscillator
strengths of Fuhr, Martin \& Wiese 1988 (NIST database), which are
generally in reasonable agreement with recent experimental values of
Bergeson et al. 1996. The only data source that has significant
outliers are the theoretical values from the Iron Project (Nahar
1995), with some discrepancies as large as a factor of 5. For one of
the important transitions for this study ($\lambda$ 2374.46 \AA, a
line from the \feii\ ground level) the Nahar value is almost 2 times
larger than the others. Using observations of \feii\ absorption lines
from the interstellar medium Cardelli \& Savage 1995 derived an
oscillator strength of 0.0326(14) for this transition, in good
agreement with the laboratory experiments of Bergeson et
al. 1996. Thus the theoretical value of 0.0527 derived by Nahar 1995
is unlikely to be correct. For this transition only, we have not used
the oscillator strength of Fuhr, Martin \& Wiese 1988 but instead a
compromise between the two recent experimental values of 0.032. More
recent theoretical calculations (Raassen \& Uylings 1998) appear to
reproduce the experimental values quite well, but are not as extensive
as the Iron Project results.

Another source of uncertainty is the exact level of the continuum.
The continuum contains several undulations, that are likely to be due
to the presence of broad \feii\ emission, as can be seen from figure
\ref{totspec}, or even more clearly from the low resolution spectrum
of FBQS1044+3656 published in White et al. 2000 which covers a larger
wavelength range. The relative contribution of the \feii\ BELs to the
continuum can be estimated from the broad trough in the spectrum
around 2650 \AA\ which is actually a gap between \feii\ emission line
complexes (see e.g., Verner et al.\ 1999).  This trough is also
present in the LBQS composite spectrum (Francis et al.\ 1991), but is
especially deep in FBQS1044+3656, indicating that at least 30\% of the
apparent continuum is actually broad \feii\ emission.  The strongest
\feii\ emission lines are expected at the same wavelengths where the
absorption lines occur (Verner et al. 1999), and since the main
blended absorption complexes themselves are also quite broad, it might
well be possible that there is emission at the position of the
absorption complexes that has not been accounted for in our continuum
normalization. The spectrum of J121442+280329 published in White et
al. 2000 is a clear example of a QSO spectrum in which a complex
interplay between \feii\ absorption and emission is shaping the
spectrum, and modelling of the absorption would be impossible without
considering the emission lines. In the case of J121442+280329 the
problem is more obvious because the emission lines are narrower. Note
that the lines from the \feii\ ground level are distributed over a
wavelength range several 100 \AA\ wide, i.e. they are separated by
much more than the width of a single \feii\ broad emission line. This
means each absorption line will be affected by a different set of
\feii\ emission lines, unlike the case of most doublet lines where the
separation is much smaller than the emission line width, and the
presence of broad emission is unlikely to have a strong differential
effect on the continuum level between the positions of the two doublet
components.
 
\subsubsection{Partial covering \label{partcov}}

The five strong lines arising from the \feii\ ground level can not be
fitted simultaneously using the preceding template approach when it is
assumed that the continuum source is completely covered by the
outflow.  Partial covering has been shown to be important in several
other QSOs with absorption lines (e.g., Hamann, Barlow, \& Junkkarinen
1997, Barlow \& Sargent 1997, Arav et al.\ 1999b), usually by
considering the relative depth of the two lines of one of the major
doublets, such as \civ\ and \siiv.  When relatively narrow absorption
components are present so that the two lines of the doublet are
visible (which implies that this type of analysis can  generally only be
applied to AALs and not BALs), it is possible to derive a model for
the outflow in which at each velocity a fraction $C(v)$ of the
continuum source is covered with a homogeneous layer of absorbing gas
with optical depth $\tau_1(v)$ in line~1 of the doublet.  This is done
by solving the set of coupled equations
\begin{eqnarray}
I_1(v)&=&(1 - C(v)) + C(v) e^{-\tau_1(v)} \\
I_2(v)&=&(1 - C(v)) + C(v) e^{-\frac{(f\lambda)_2}{(f\lambda)_1}\tau_1(v)}, 
\label{coveq}
\end{eqnarray} 
for every velocity $v$.  Here $I$ is the residual intensity in the
normalized spectrum, the subscripts 1 and 2 refer to the two lines of
the doublet, and $(f\lambda)_2 / (f\lambda)_1$ is the known ratio of
the optical depths in two lines that originate from the same lower level.

The same analysis can be applied to every pair chosen from the 5 lines
arising from the \feii\ ground level, and if the partial covering
model used is correct, then each pair should yield the same covering
factor as a function of velocity, providing a strong confirmation of
the simple partial covering model.  Several complications crop up when
analysing the \feii\ lines in this way, the most serious being line
blending.  If one of the two lines used to solve for optical depth and
covering factor is contaminated by another absorption line, the
solution will be erroneous and often no solution is possible at all.
A second problem is that the oscillator strengths of the lines are not
accurately known, in contrast to the case of the doublet lines where
the ratio is exactly 2.  When the optical depth is very low, small
continuum errors also cause problems.

In spite of these difficulties, we can still derive some insights from
such a covering factor analysis.  We first consider the $C(v)$ that
can be derived from the two \feii\ lines at 2600~\AA\ and 2586~\AA;
these suffer the least from blending with other lines and therefore
give the cleanest solutions.  In figure \ref{covfac}, we show what
fraction of the continuum source must be uncovered to make the
profiles of these two lines consistent, both for cluster 1 and cluster
2.  At some velocities, contamination by other lines or noise do not
allow a solution for $C(v)$, and we have indicated these points with
asterisks above the spectrum.  To reduce the effect of noise, the
spectrum was smoothed with a 5-point moving average before solving
equation \ref{coveq}.
The shape of the strongest line is almost entirely determined by the
covering factor, as is expected on the basis of our previous result
that the template fitting procedure predicts a much smaller flux at
the bottom of the strongest \feii\ lines than is actually observed.
This applies both in cluster 1 and cluster 2.  Thus, the residual flux
at the bottom of the strong lines is either due to a small part the
continuum source not being completely covered by the outflow, or due to 
scattering of continuum light around the absorbers. In the case of
FBQS1044+3656 we consider the first possibility more likely, because 
the strong variations of covering factor as a function of velocity are 
not expected if the remaining flux is due to scattered light.

The covering factor solutions based on other line pairs suffer more
from all of the problems mentioned above. Nevertheless, it is very
important to check whether the solutions obtained from different pairs
yield more or less consistent covering factors.  Cluster 1 is too much
affected by blending to make a useful comparison.  In figure
\ref{covcomp}, we compare the solutions for the covering factor in
cluster 2 for the 5 most suitable line pairs.  In view of all the
uncertainties, the solutions are consistent.  There is a systematic
difference between the pairs including the line 2344.2 \AA\ and the
other pairs, in the sense that they yield a smaller covering factor
outside the deepest components in the profile.  Such behavior could be
caused by an error in the oscillator strength of the line 2344.2 \AA,
an error in the continuum level, or by a wavelength dependent
continuum source size.  Since the line is close to the blue cut-off of
the spectropgraph, the S/N ratio at the position of the 2344.2 \AA\
line is considerably worse than at the position of the other lines,
which may also contribute to the discrepancy. Since our continuum
consists of a combination of a smooth underlying continuum generated
in a very small region and broad undulating \feii\ emission lines from
a much larger region, we consider a wavelength dependent continuum
source geometry to be the most likely explanation.

In principle, the covering factor solution also yields the optical
depth as a function of velocity, and hence the column density.  The
velocity integrated column densities derived in this way for each
cluster are typically 2-3 times higher than those derived from the
template fitting.  Since the column densities derived from template
fitting mainly depend on the weakest non-saturated lines, this
difference can not be attributed to the effects of partial
covering. Instead, the cause of the discrepancy is that the integrals
of optical depth over velocity are dominated by large peaks that occur
when the residual fluxes in the two lines used to derive the covering
factor solution become almost equal, which can occur because of small
continuum errors when the lines are weak, or because of another
absorption line falling into the profile of the weaker line.  This
forces a solution with very high optical depth at this velocity.
Since this is not physical, we consider the column densities derived
from the template fitting more reliable.

We now verify that the results for the level populations obtained by
the covering factor approach and the template fitting procedure 
are consistent.
The foregoing analysis shows that the covering factor completely
determines the profiles of the two strongest \feii\
lines, so the covering
factor as a function of velocity can be estimated from the remaining
flux in these lines as,
\begin{equation}
(1 - C(v))= {1 \over 2} [I_{2382}(v) + I_{2600}(v)] \label{covmean}.
\end{equation}
The thick line in Figure \ref{fe2plot} again shows the model fit based
on the parameters in Table 2, this time including the effect of the
covering factor defined by equation \ref{covmean}.  All the lines are
now well-reproduced by the model. Note that if multiple lines from gas
at different velocities (and hence with different covering factors)
contribute to the absorption at a given wavelength, the total
remaining flux depends on whether the gas at different velocities
covers the same region of the background source or not. For the model
presented here, we have assumed that there is no correlation between
the positions of the covered or uncovered regions for gas with
different velocities. Numerically, this means that each line
contributing to the absorption at a given wavelength reduces the flux
by a factor $r$ equal to
\begin{equation}
r=(1 - C(v)) + C(v) e^{-\tau (v)}
\label{covcor}
\end{equation}   

\subsection{Other lines}

The only other feature that we can identify with certainty is \mnii\
absorption from cluster 2.  All three lines from the \mnii\ ground
state (2576.877, 2594.499 and 2606.462 \AA) are detected.  The depth
of the \mnii\ features is well fit if we use the \mgi\ absorption
template, and assume that the column density of \mnii\ is $10^{-2}$
times the column density of \feii\ ($\log N_{Mn~II}= 13.3$), in good
agreement with standard abundance ratios. In Figure \ref{fe2plot} the
positions of the \mnii\ absorption lines are indicated.

Absorption lines of \fei\ are absent from the spectrum, providing an
important constraint on the ionization parameter of the absorbing gas.
The strongest line from the \fei\ ground level (2484.0209 \AA\ rest
wavelength) is not detected (see figure \ref{fe2plot}).  We have
determined an upper limit on the \fei\ column density by trying to fit
model \fei\ spectra using the \mgi\ absorption template, and
estimating by eye how much \fei\ could be tolerated without the line
being clearly detected. Because of the relatively smooth continuum in
the relevant part of the spectrum, we are able to constrain the column
density of \fei\ in the ground state to $\log N_{Fe~I} < 13.0$.  It
would be useful to have some information on the \feiii\ abundance
(WCP), but our wavelength coverage does not include any \feiii\ lines.

\section{DISCUSSION}

\subsection{The electron density in the outflow}

Our most important result is the finding that absorption from excited
levels of \feii\ is detected, and that the level populations are
smaller than expected for LTE.  This shows that the electron density
in the outflowing clouds in FBQS1044 is lower than that in the
lower velocity clouds in spectrum of Q0059-2735 studied by WCP, which
were found to be in LTE. An accurate electron density estimate based
on the observed level populations is difficult to obtain, since only a
small fraction of the radiative transition probabilities for the
strongly forbidden transitions within and between the $a^6D$, $a^4F$
and $a^4D$ terms are known (Nussbaumer \& Storey 1980, Quinet, Le
Dourneuf \& Zeippen 1996).

A very rough estimate of the electron density can be obtained by
considering only the relative populations of the ground level and the
lowest excited level ($E=384\ \rm{cm}^{-1}$). This level can only
decay radiatively to the ground level and not make other radiative
decays with unknown transition probabilities. The electron density can
be estimated assuming that there is equilibrium between collisional
excitation into this level from the ground level and radiative
de-excitation to the ground level. For this transition the radiative
decay probability $A_{21}$ is 0.00213 (Quinet, Le Dourneuf \& Zeippen
1996) , and collision strength $\Omega_{12}$ is 5.5 (Zhang \& Pradhan
1995). Taking into account the formal errors on the level populations
in Table 2, and constraining the temperature of the absorbing gas to
lie in the range between 300 and 10.000 K, we obtain $\log n_e =
3.55(0.25)$ for cluster 2 and $\log n_e = 3.0(0.4)$ for cluster
1. Using the observed level populations and the collision strengths
from Zhang \& Pradhan 1995, it is easily verified that for $\log n_e =
3.55$ and a temperature of 5000 K the total collisional excitation
rate out of the lowest excited level is 0.24 times the radiative decay
rate to the ground level, so that its neglect is justified for this
rough estimate.

We can also derive an estimate of the electron density by using the
most recent attempt at modelling the level populations of \feii\ as a
function of density using a full multilevel calculation by Verner et
al. 1999. In these models the unknown transition probabilities have
been set to such a small value that they are effectively
neglected. Based on these results we can make an estimate of the
electron density by comparing the departure coefficient of the ground
level (the only level among the ones relevant here for which results
are presented in Verner et al. 1999) with the predicted value as a
function of electron density. We define an ``observed'' departure
coefficient for the ground level as
\begin{equation}
b_{0,obs}=\left({n_{0,obs}\over{\sum n_{i,obs}}}\right)
\left({{\sum g_i e^{(-E_i/kT)}}\over{g_0}}\right)
\end{equation}
where the sum is over the levels $i$ given in table \ref{levpop}, and
the temperature was taken to be 5000 K for consistency with the Verner
et al. model. For this temperature, the contribution of the
higher energy levels to the partition function will be very small
because of the Boltzmann factor, and their neglect is well
justified. The observed departure coefficients are $\log b_{0,obs} =
0.33(0.08)$ for cluster 2 and $\log b_{0,obs} = 0.37(0.17)$ for
cluster 1. Using Figure 3 in Verner et al. 1999 we can convert these
limits to $\log n_e = 3.8(0.3)$ for cluster 2 and $\log n_e =
3.5(0.5)$ for cluster 1. These values are in reasonable agreement with
the results obtained above by treating the ground and lowest excited
level of \feii\ as a two-level atom.

\subsection{Ionization modelling}

We have used the photoionization code CLOUDY (Ferland 1996) to model
the ionization state of the outflow to derive additional constraints
on the absorbing gas.  Our observational data yield the following
constraints on the outflow in FBQS1044+3656: a \mgi\ column density
$\sim 2 \times 10^{13} {\rm cm^{-2}}$, a lower limit on the \mgii\ /
\mgi\ ratio of $\sim 30$, an \feii\ column density $\sim 3 \times
10^{15} {\rm cm^{-2}}$, an upper limit on the \fei\ column density of
$\sim 10^{13} {\rm cm^{-2}}$ and an estimate of the electron density
of $\sim 4 \times 10^3 {\rm cm^{-3}}$.  It turns out that the lower
limit on the \mgii\ / \mgi\ ratio is not very useful, so we will
concentrate on the \mgi, \feii, and \fei\ column densities.

\subsubsection{Dust-free models }

We will first consider simple models of the absorbing clouds,
investigating their structure as a function of ionization parameter,
density and shape of the ionizing spectrum. For the moment we ignore
the possible effects of dust, which can introduce significant
complications.  The clouds are modelled as thin constant density
slabs.  In order to satisfy the observational constraints, the slabs
are truncated when a value of the \feii\ column density of $\log
N_{Fe~II}=15.5$ is reached. Furthermore, to satisfy the electron
density constraint we have adjusted the density of the slab in each
model so that the \feii\ weighted mean electron density is $4 \times
10^3 {\rm cm^{-3}}$. In this way the ionization parameter , the
abundances and the shape of the ionizing spectrum remain as
parameters that can be adjusted to satisfy the constraints on the
\mgi\ and \fei\ column densities. The parameters and properties of the
ionization models presented here are given in Table \ref{ionmods}.

Figure \ref{uplot} illustrates the structure of the slab as a function
of ionization parameter for gas with solar composition irradiated by a
standard Matthews \& Ferland (1987) AGN spectrum.  When modelling the
very low ionization species considered here with the CLOUDY code, one
encounters the difficulty that \mgi, \fei\ and \feii\ can occur in gas
with a temperature below the limit at which CLOUDY results are
reliable.  We therefore had to stop the calculation of the cloud
structure at the depth where the temperature falls below 1000~K, if
this occurs before the required \feii\ column density is reached.  At
temperatures below 1000~K \feii\ is still the main form of Fe.  This
means that when the calculation is stopped because of the temperature
constraint (which occurs for low ionization parameters), the column
densities found are only lower limits.

The run of temperature and electron density through the slab is
similar to that found in the detailed \feii\ BEL models of e.g. Wills
et al. (1985) and Collin-Souffrin et al. (1986), with an almost fully
ionized Str\"omgren layer at the illuminated side of the side,
followed by a partially ionized zone in which most of the column
density of \feii\ resides. To understand the behaviour of the \mgi\
column density as a function of the model parameters it is important
to realize that the ionization equilibrium of Mg behaves slightly
counter-intuitively, in the sense that the fraction of neutral Mg
drops sharply {\em behind} the hydrogen ionization front. The reason
for this behaviour is that the photoionization rate of \mgi\ is hardly
affected by the front because the ionization potential of \mgi\ is
considerably lower than that of hydrogen, whereas the recombination
rate is much reduced because of the sharp drop in electron
density. Therefore, the \mgi\ column density of the slab resides
almost entirely in the ionized Str\"omgren layer.

The three models with $\log U \ge -3$ all have \mgi\ column densities
less than $10^{13}{\rm cm^{-2}}$ (see Table \ref{ionmods}), between 3
and 10 times less than observed in FBQS1044+3656. The three models
with $\log U$ equal to --4, --5 and --6 run into the 1000~K
temperature limit before a high enough \feii\ column density is
reached, so that the column densities of \fei\ and \mgi\ are lower
limits only. The model with $\log U=-6$ is excluded, since it already
has an \fei\ column density higher than our upper limit when the
\feii\ column density is only $2.5~10^{14}{\rm cm^{-2}}$. Thus of the
models presented in Figure \ref{uplot}, only those with $\log U = -4$
and $\log U = -5$ may explain the observed column densities in
FBQS1044+3656. Although the column densities may be consistent with
such a very low ionization parameter model, they are not likely to
provide a good explanation for the observed \feii\ lines in detail,
since most of the \feii\ column density would have to reside in the
region with a temperature below 1000~K, which is difficult to
reconcile with the observed population of the excited state at 1873
cm$^{-1}$ because the Boltzmann factor for this level should be $ <
0.067$. Thus we conclude that simple solar composition ionization
models are not able to reproduce the observed column densities.

Model 2 ($\log U = -2$) only fails to reach the observed \mgi\ column
density by a factor of 3. Such a relatively small discrepancy may be
fixed by changing the abundances from the solar values. The simplest
solution is to increase the Mg abundance by a factor of 3, which
brings the \mgi\ column density very close to the observed
value. Increasing the Fe abundance reduces the thickness of the slab
for a given \feii\ column density, but because most of the \mgi\
resides in the Str\"omgren layer, this hardly affects the \mgi\ column
density. Thus with an overall increase in metallicity to 3 times solar
(Model 7) the observed column densities can be reproduced.

Although the column densities can be reproduced by this high
metallicity model, there are two problems that remain to be
resolved. The first is that these models are rather contrived in the
sense that extreme fine tuning of the cloud geometry is required to
obtain the correct \feii\ column density. Most \feii\ resides in a thin
layer at the back side of the cloud, with a thickness much smaller
than the thickness of the Str\"omgren layer, a problem also pointed
out by WCP (see below).  Since the thickness of the slab is set by the
\feii\ column density, this problem can be remedied by reducing the Fe
abundance by a very significant factor. This leads to slabs with a
much thicker partially ionized zone. The larger extent of the
partially ionized zone does not contribute appreciably to the \mgi\
column density however, so that an enhanced Mg abundance is still
required. The second problem is that the \mgi\ and \feii\ reside in
different zones of the slab. In FBQS1044+3656, the gas along the line
of sight is broken up into many clouds at different velocities, and we
would not expect to see such a good correspondence between the
absorption components of \feii\ and \mgi\ if these lines were not
formed in the same gas.

In figure \ref{spplot} we investigate the sensitivity of these results
to our assumptions about the shape of the ionizing spectrum. The AGN
spectrum used above has a very strong EUV bump. Some more recent
observation (e.g. Laor et al. 1997) suggest that in at least some
active galaxies the EUV flux is considerably smaller than predicted by
the Matthews \& Ferland spectrum. To investigate this effect, we have
repeated some models with a simple power law spectrum with slope -1.4
between 10 microns and 50 keV, which effectively removes the EUV
bump. Figure \ref{spplot} compares the models with $\log U$ equal to
-2, -4 and -6 for the two choices of ionizing spectrum. Although the
temperature and electron density in the partially ionized zone are
higher in the case of the power law spectrum, the column densities are
very similar, so that our conclusions from Figure \ref{uplot} are not
very dependent on the exact shape of the ionizing spectrum.
 
\subsubsection{Models with dust}

From the ionization models above we have found that the low ionization
species we have been studying are mainly found in gas with a
temperature of a few thousand Kelvin. This means it is very well
possible that some form of dust will be present in the outflowing
clouds. As we will demonstrate, this can change the ionization models
significantly. The presence of dust will have three main effects:

\begin{enumerate}
\item Fe is likely to be strongly depleted. This means
that the hydrogen column density required to obtain the observed
\feii\ column density is much larger.

\item It is possible that Fe and Mg are depleted by {\em different}
depletion factors. Thus the column density ratios between Fe and Mg
ions will be changed.

\item The dust absorbs ionizing radiation, which changes the
ionization equilibrium in the slab.
\end{enumerate}

The following models are based on the Orion dust model in CLOUDY,
which is in turn based on the work on abundances and dust properties
in the Orion Nebula H~II regions (Ferland 1996, Baldwin et al. 1991,
Osterbrock et al. 1992). The properties of this dust model that are
important for this work are the depletions of Fe and Mg ($\sim$ 0.1
for both) and the dust opacity, which reflects the absence of small
grains. The physical conditions in the clouds we consider are more
similar to those in H~II regions than those in the local ISM, for
which a different depletion pattern and grain size distribution has
been derived(e.g Cowie \& Songaila 1986).

The results of the ionization models including dust are illustrated in
Figure \ref{udust}. The most obvious effect of the presence of dust is
the depletion of Fe, which leads to thicker slab models with a much
thicker partially ionized zone, as discussed above. However, the most
important physical difference between these models and the models
without dust is that the \mgi\ fraction no longer drops as steeply
behind the ionization front, so that the largest contribution to the
\mgi\ column density now comes from the partially ionized zone rather
than the Str\"omgren layer. The reason for this behaviour is that the
\mgi\ ionizing flux (including trapped Ly $\alpha$) is absorbed by
dust, so that the reduction in recombination rate due to the drop in
electron density is partially compensated. Thus in the models with
dust the \feii\ and \mgi\ column density reside in the same region of
the slab, whereas they are separated in the models without dust. Since
the partially ionized zone now contributes to the \mgi\ column, the
observed column can be explained with a much smaller Mg abundance than
in the models without dust.

The largest \mgi\ column density for the given \feii\ column density
is produced for $\log U = -3$ (Model 13).  Even for this model it is
still necessary to assume that Mg is three times less depleted than in
the standard Orion dust model (i.e. by a factor of 0.33 instead of
0.1) to reproduce the observed \mgi\ column. For lower ionization
parameters the models run into the 1000 K temperature limit again,
which means that most \feii\ would be too cold to explain the \feii\
excitation, as discussed above. Higher ionization parameters can also
reproduce the \mgi\ column density if we allow Mg to be even less
depleted.  However, the models with larger ionization parameters are
effectively excluded by another constraint, which is that the spectrum
of FBQS1044+3656 does not appear to suffer from very strong
extinction. The extinction at 2500 \AA\ predicted by the models with
dust is given in Table \ref{ionmods}. A comparison of the spectrum of
FBQS1044+3656 with the mean QSO template of Weymann et al. 1991 shows
that the extinction at 2500 \AA\ is not likely to be larger than 1.0
magnitude, unless the spectrum is intrinsically very steeply rising
towards shorter wavelengths. This excludes the models with higher
ionization parameters than $\log U = -3$.   

All models with $\log U > -6$ have \fei\ column densities that lie
well below our upper limit.  We conclude that models with dust can
explain the observed properties of the outflow reasonably well, and do
not suffer from the problems with an improbable cloud geometry and
separation of the \mgi\ and \feii\ producing zones encountered in the
high metallicity models without dust.

\subsection{The size of the outflow}

For the cosmological parameters $H_0 = 50\ {\rm km\sec^{-1}
Mpc^{-1}}$, $\Omega=1$ and $\Lambda=0$, the absolute B magnitude of
FBQS1044+3656 is $M_B=-26.2$. Assuming further that the shape of the
continuum is close to the typical AGN spectrum described by Matthews
\& Ferland (1987), we estimate the luminosity of FBQS1044+3656 to be
close to the canonical QSO value of $10^{46} {\rm erg\ s^{-1}}$.  In
table \ref{ionmods}, the distance between the source of ionizing
radiation and the absorbing cloud corresponding to each model is
tabulated assuming this luminosity. Both the high metallicity models
without dust and the dusty models require that the outflow is situated
at a distance of several hundred parsecs from the source of ionizing
radiation, i.e typical of the QSO narrow line region. This distance is
about two orders of magnitude larger than the distance derived by WCP
for the \feii\ absorbing clouds in the BALQSO Q0059-2735.

This large size of the outflow raises several questions. The velocity
of the outflow is considerably larger than that of material in the
narrow line region, and is more typical of the BEL clouds that are
situated at only 0.1 parsec from the nucleus. The combination of a
large distance and the very small cloud size implied by partial
covering re-emphasizes the problems associated with the stability of
small, high speed clouds with a very low filling factor , which have
been discussed extensively in the context of the formation of broad
absorption lines (Weymann, Turnshek \& Christiansen 1985, Begelman, de
Kool \& Sikora 1991, Weymann 1997). Although the small implied sizes
of the clouds are hard to understand theoretically, we do point out
that the small transverse cloud size implied by partial covering is
consistent with the physical thickness of $\sim 10^{16}$\ cm of the
slabs in our ionization models as derived from the hydrogen densities
and column densities.
  
One of the main objections to large distances for BAL-like outflows
comes from constraints on the mass and energy involved using the
typical parameters of high-ionization BALs.  The low ionization
absorption lines studied in this paper, however, have rather low
velocity and equivalent width compared to typical high-ionization
BALs.  We can estimate the mass loss rate and kinetic energy of the
outflow for the specific case of FBQS1044+3656 to see if they are
unreasonably large.  The mass involved in the low ionization
outflow can be estimated from
\begin{equation}
M_{outflow} \gtorder 10^7 N_{21} R_{21}^2 f_{cov} \Msun \quad .
\end{equation}
where $N_{21}$ is the hydrogen column density normalized to $10^{21}$\
cm$^{-2}$, $R_{21}$ the size of the outflow in units of $10^{21}$\ cm
and $f_{cov}$ is the global covering factor of the outflow.  Since \feii\
absorption is not very common in QSOs, $f_{cov}$ could be as small as
$10^{-2}$ for \feii\ absorption-producing outflows. Alternatively the
outflows may be short lived phenomena, in which case the constraints
on the mass loss rate are also less severe.  Taking the parameters of
our preferred Models 7 or 15 and estimating the flow time to be
$R/v$, the implied mass outflow rate is $70 f_{cov}$ \Msun yr$^{-1}$
and the kinetic power of the outflow is $3 \times 10^{44} f_{cov}$ erg
s$^{-1}$.  These are values that do not argue strongly against the
large size of the outflow.

\subsection{Is FBQS1044+3656 a BALQSO?}

As mentioned in the Introduction, the characteristics of the
absorption lines in FBQS1044+3656 are intermediate between those of
BALs and AALs.  Weymann et al. (1991) classified a QSO as a BALQSO if
the \civ\ line profile showed a continuous absorption of at least 10\%
in depth spanning more than 2000 \kms\ in velocity, discounting
absorption closer than 3000 \kms\ bluewards of the emission peak.  The
\mgii\ absorption in FBQS1044+3656 meets these criteria, even when a
correction is made for the fact that the doublet separation in \mgii\
is larger than in \civ\ .  The total width of the \mgii\ absorption is
about the same as in Q0059-2735 which exhibits extremely strong BALs
in the higher ionization lines, and it is an established property of
low ionization BALQSOs that the absorption troughs of higher
ionization species like \civ\ or \ovi\ are generally much 
wider than the \mgii\ absorption (Weymann et al. 1991).  Thus, 
considering the properties of the \mgii\ absorption alone FBQS1044+3656
should be classified as a BALQSO. In the standard picture of BALQSOs,
the absorption lines are formed in an outflow with a typical size of a
few parsec.

The characteristics of the very low ionization lines of \mgi\ and
\feii\ are typical of AALs (Weymann et al. 1979, Foltz et al.  1986),
for which no firm distance scale has been established so that the
distance of a few hundred parsec found in this paper is perhaps not a
surprise. The remarkable fact lies in the excellent correspondence in
velocity between the strongest \mgii\ features and the \mgi\ and
\feii\ absorption lines, which establishes without a doubt that the
very low ionization lines and the \mgii\ lines are formed in the same
outflow.  The association of the \feii\ and \mgi\ absorption with the
\mgii\ BAL is much clearer in FBQS1044+3656 than in Q0059-2735, which
only shows a number of relatively narrow \feii\ components with
outflow speeds of a few hundred kilometer per second, and a very
smooth \mgii\ BAL profile.

The combined properties of the \mgii\ , \mgi\ and \feii\ lines seem to
indicate that FBQS1044+3656 contains an outflow with BAL-like
characteristics situated at a distance of a few hundred parsec from
the nucleus, a scale very much larger than would have been attributed
to the outflow on the basis of the \mgii\ lines only. At this time,
FBQS1044+3656 is a unique object and we do not know how general these
results are.  In view of the importance of establishing the relation
between the large scale outflow found here and the classical high
ionization BAL phenomenon, the essential next step in the study of
FBQS1044+3656 is to obtain a UV spectrum of the source to compare the
absorption profiles of the low and high ionization species.

\section*{ACKNOWLEDGMENTS}

Part of this work was performed under the auspices of the
U.S. Department of Energy by University of California Lawrence
Livermore National Laboratory under contract No.~W-7405-Eng-48.

\newpage

\figcaption[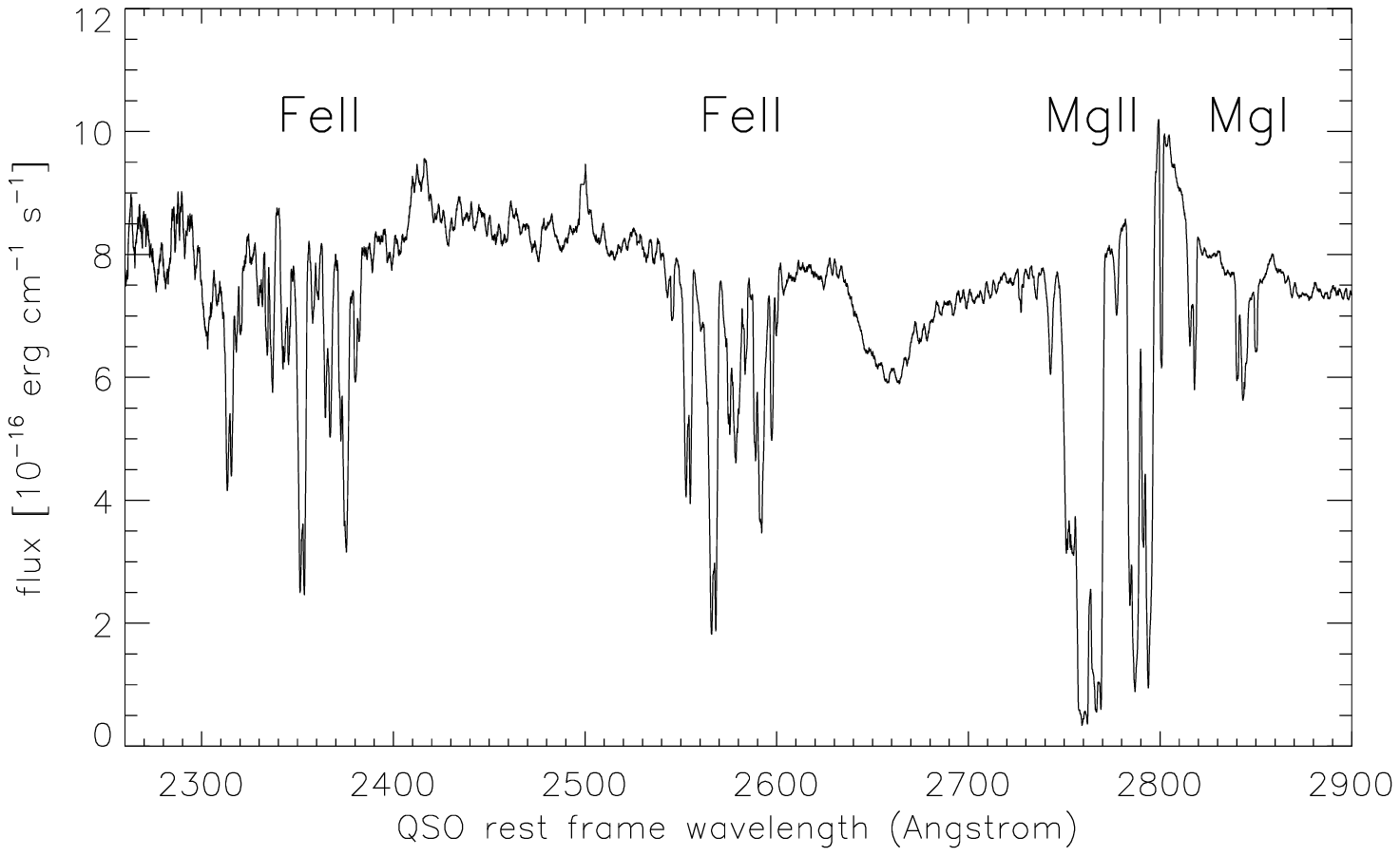]{A smoothed plot of the spectrum of
FBQS1044+3656, giving an overview of the regions analyzed in this
work. \label{totspec}}

\figcaption[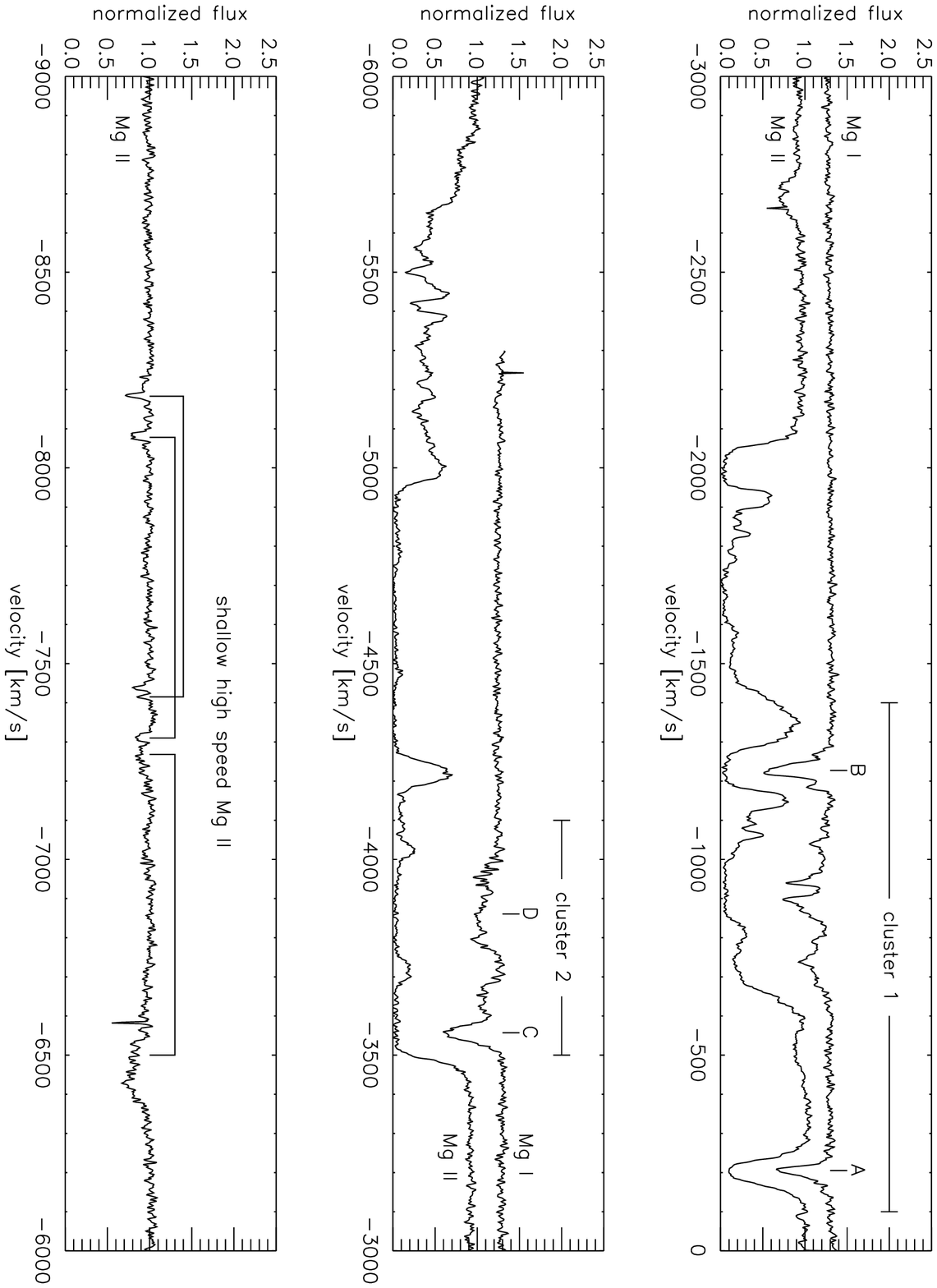]{ The absorption lines of \mgi\ and \mgii\ as a
function of velocity. The spectrum on the \mgi\ velocity scale has
been offset by 0.3 for clarity. The \mgii\ velocity scale is based on
the red component of the \mgii\ doublet.  The line complexes discussed
in the text are labelled. \label{mgplot}}

\figcaption[fig3.eps]{ The spectrum of FBQS1044+3656 in the region
of the \feii\ lines. The thin line represents the model fit based on
the template fitting method and the parameters in Table \ref{levpop}
assuming the background continuum source is completely covered by the
outflow. The model shown as the thick line is based on the same model
parameters, but assumes that the covering factor as a function of
velocity is given by the absorption profile of the strongest \feii\
lines, as described in the text. The expected positions of the \feii\
lines caused by the gas responsible for absorption components A,B,C
and D of the \mgi\ profile (defined in Figure \ref{mgplot}) are
indicated above the spectrum. Lines of \mnii\ and the expected line
positions for the strongest \fei\ lines are also indicated.
\label{fe2plot}}

\figcaption[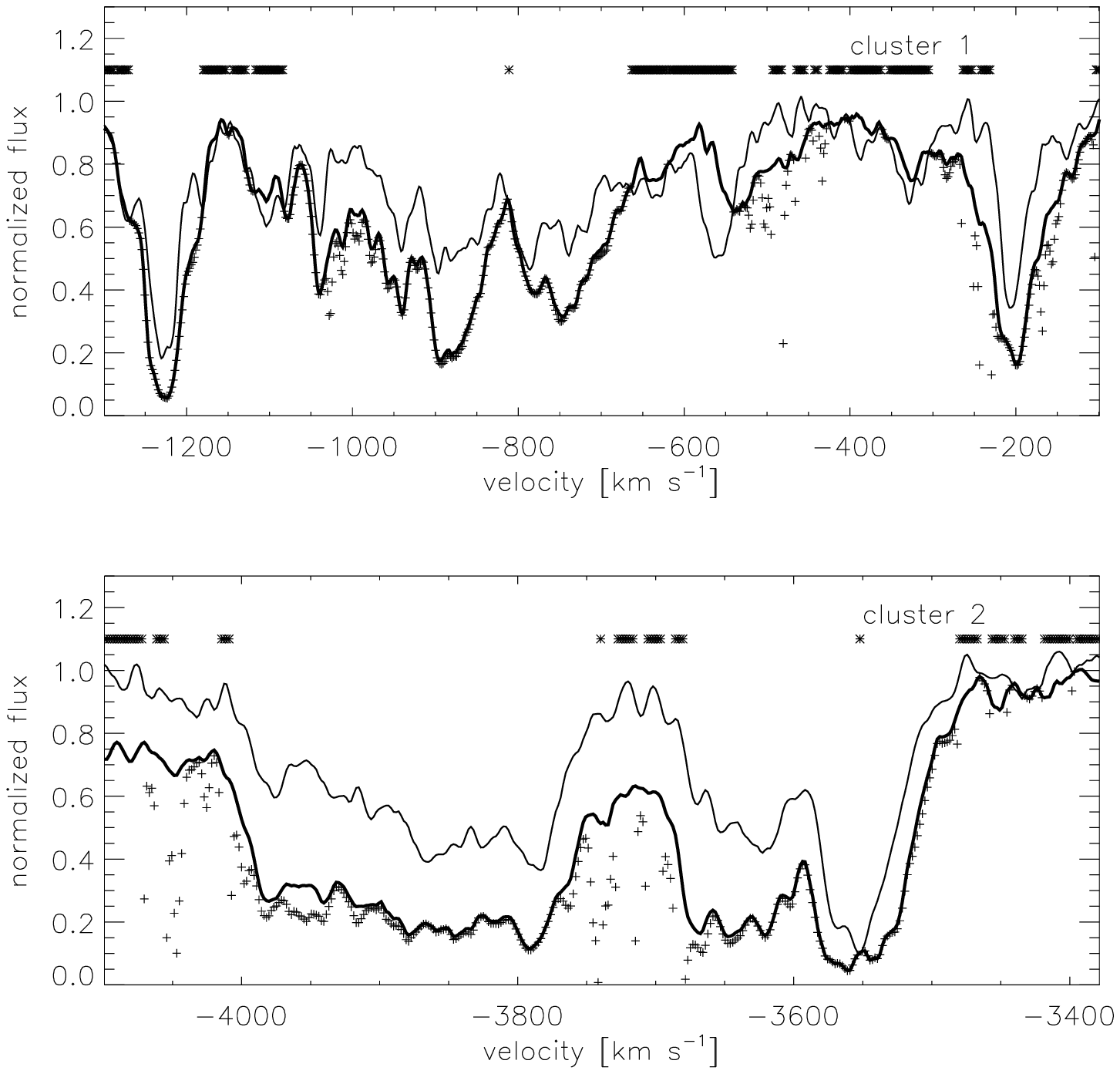]{ The covering factor derived from a comparison
of \feii\ lines $\lambda 2600$ and $\lambda 2586$. The crosses
represent the unobscured continuum flux (1 - C).  The thin and thick
lines are the profiles of the two lines used to obtain the solution
for the covering factor.  Contamination by other lines, noise and
continuum errors allow no physical solution for the covering factor at
some velocities.  The positions at which this happens are indicated by
an asterisk above the spectrum.  \label{covfac}}

\figcaption[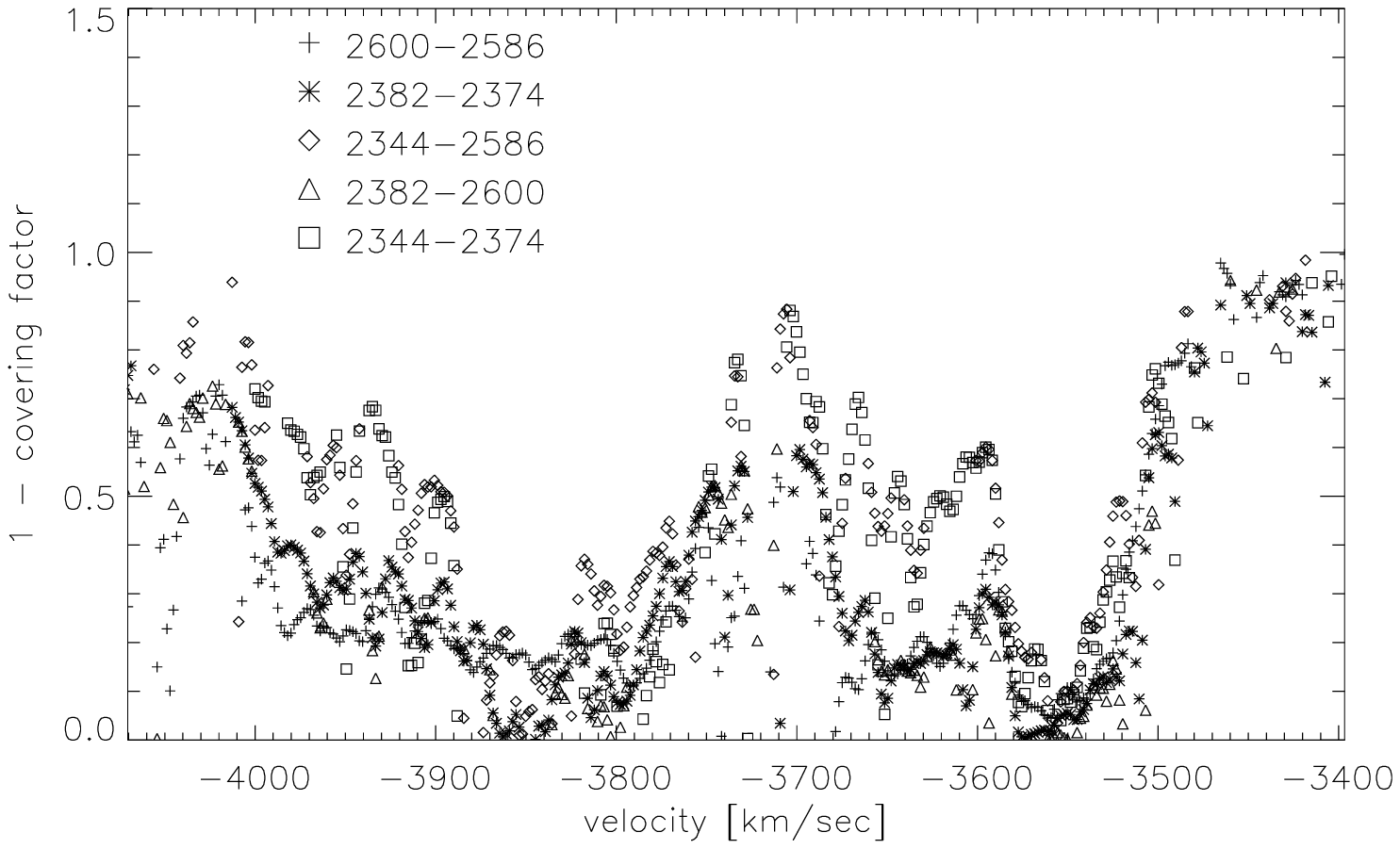]{Comparison of the solutions for the
unobscured flux in cluster 2 for 5 different pairs of lines from the
\feii\ ground state.  Points with no physical solution have been
omitted.  \label{covcomp}}

\figcaption[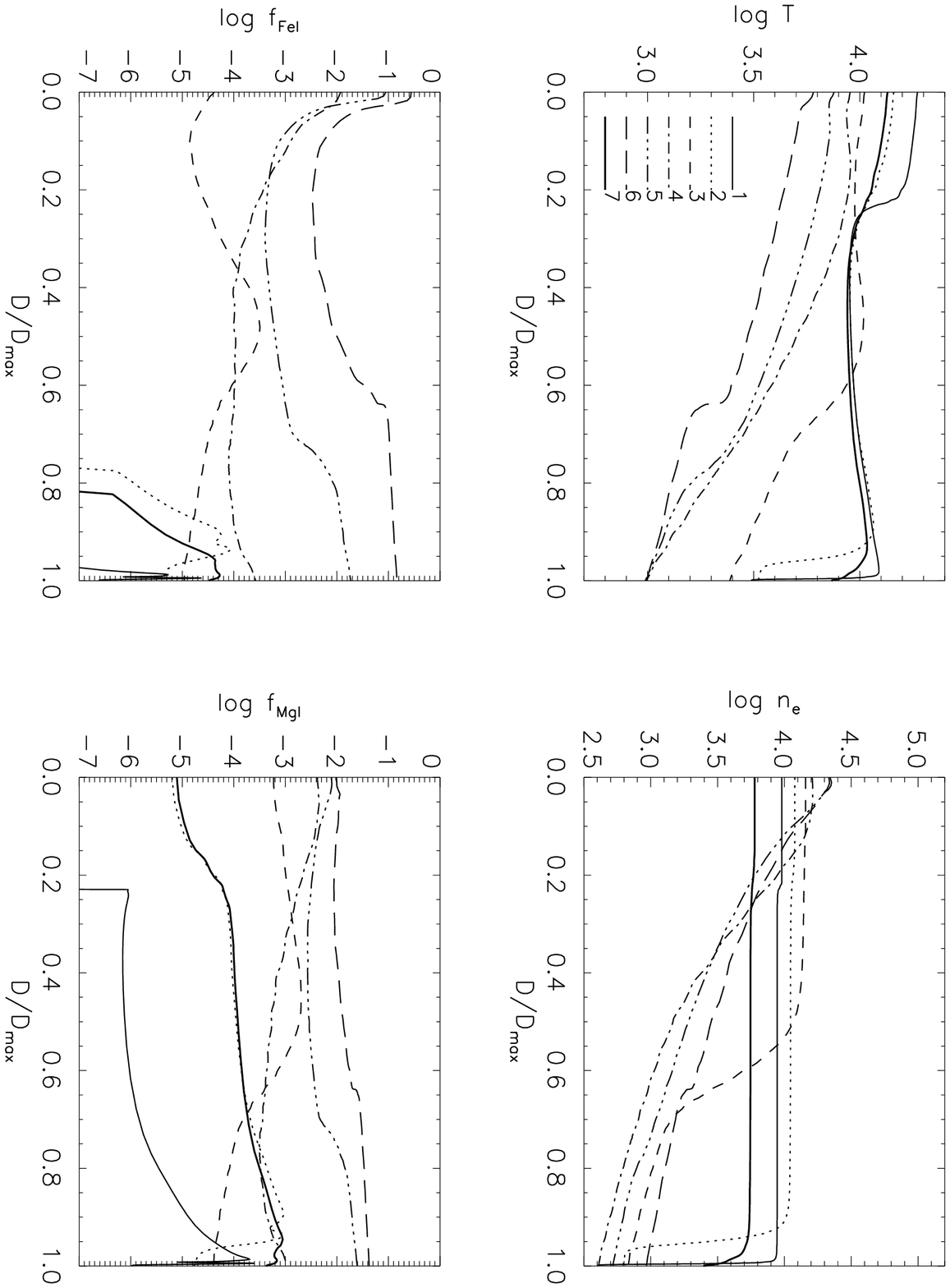]{ The structure of an irradiated slab with solar
abundances as a function of ionization parameter. The variables are
plotted as a function of position in the slab, normalized to the total
thickness of the slab $D_{max}$. The key next to the line type
definition in panel (a) refers to the models in Table \ref{ionmods}.
Panel (a) shows the electron temperature, panel (b) the electron
density and panels (c) and (d) the ionization fraction of \fei\ and
\mgi\ respectively.
\label{uplot}}

\figcaption[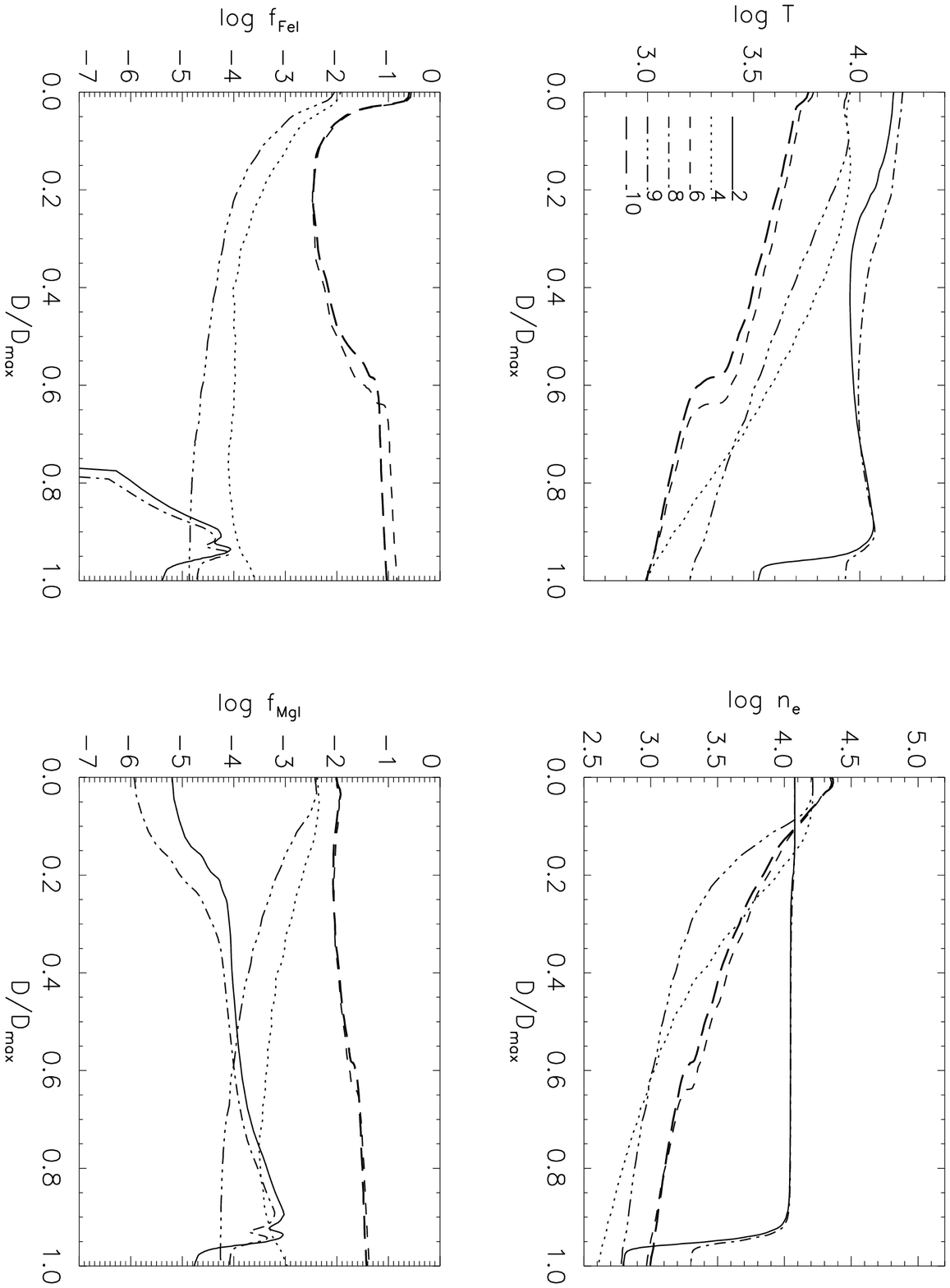]{ This figure illustrates how the ionization
models in Figure \ref{uplot} change if the spectrum of the incident
radiation is a simple power law with slope --1.4 rather than the
Matthews \& Ferland (1987) AGN spectrum.
\label{spplot}}

\figcaption[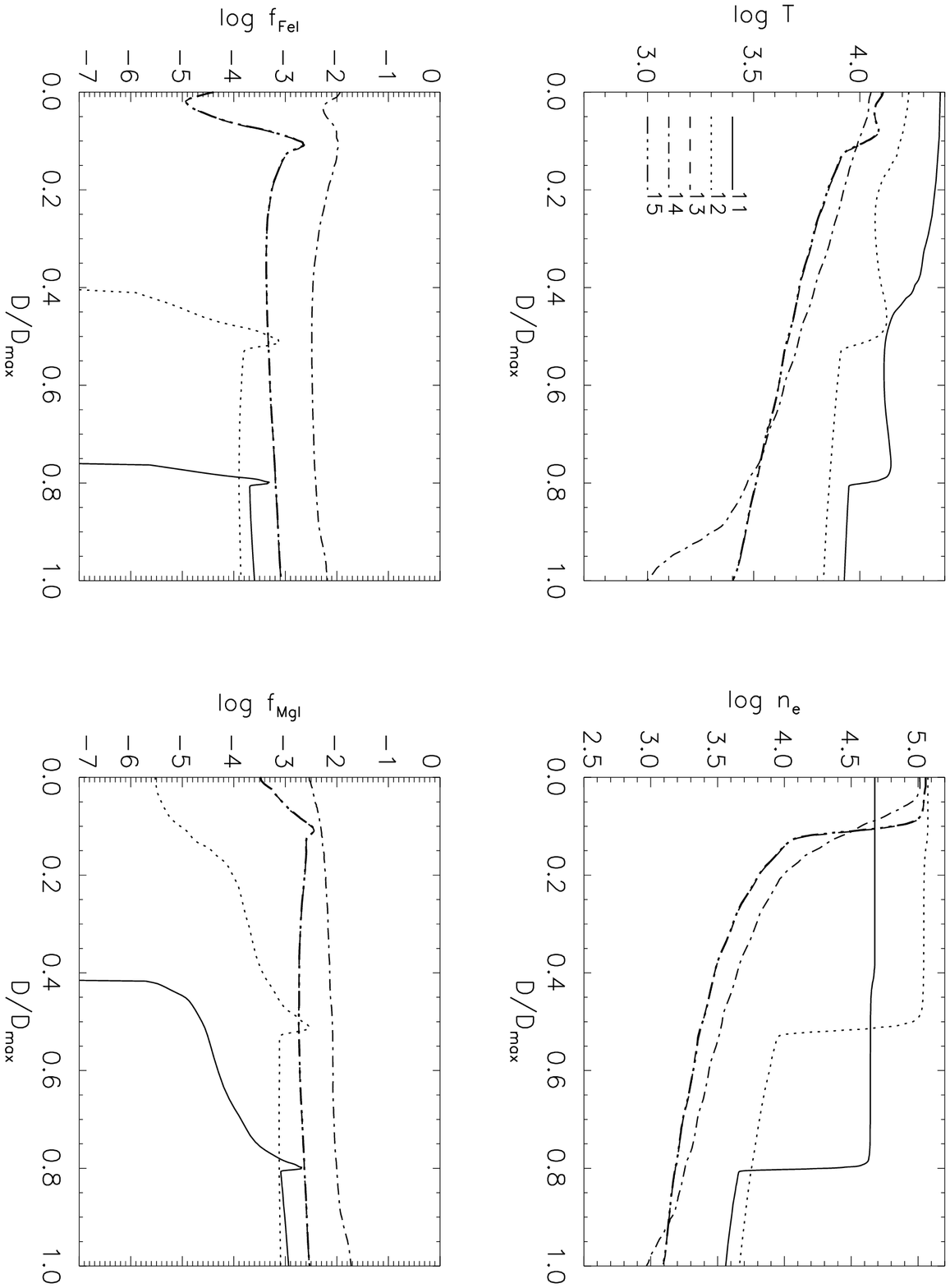]{ Photoionized slab models assuming the gas
contains dust. For the models with dust (Table \ref{ionmods}) the
elemental abundances in the gas phase are depleted as observed in the
Orion nebula, except model 15 which has a lower Mg depletion. Since
the higher Mg abundance has negligible influence on the slab structure
it is indistinguishable from model 13 in these plots. Model 15
fulfills all observational constraints including those on the
intrinsic extinction.
\label{udust}}

\begin{deluxetable}{rrrrrrrrr}
\tablecaption{\sc \feii\ lines and oscillator strengths} 
\tabletypesize{\footnotesize}
\tablehead{
\colhead{Term}
&\colhead{$\lambda_{vac}$ [\AA]}
&\colhead{E$_{l}$[cm$^{-1}$]}
&\colhead{g$_{l}$}
&\colhead{f$_{lu}$}\tablenotemark{a}
&\colhead{f$_{lu}$}\tablenotemark{b}
&\colhead{f$_{lu}$}\tablenotemark{c}
&\colhead{f$_{lu}$}\tablenotemark{d}
&\colhead{f$_{lu}$}\tablenotemark{e}
}
\startdata
$a^6D$ &   2632.10&   667.683&    6&   0.084&   0.125&   0.076&   0.085&   0.084\\ 
&   2631.83&   862.613&    4&   0.120&   0.175&   0.112&   0.131&   0.125\\ 
&   2629.07&   977.053&    2&   0.180&   0.227&   0.173&   0.173&   0.173\\ 
&   2626.45&   384.790&    8&   0.043&   0.068&   0.043&   0.044&   0.043\\ 
&   2622.45&   977.053&    2&   0.050&   0.065&   0.056&   0.056&   0.055\\ 
&   2621.19&   862.613&    4&   0.004&   0.003&   0.004&   0.004&   0.004\\ 
&   2618.39&   667.683&    6&   0.045&   0.050&   0.045&   0.050&   0.048\\ 
&   2614.60&   862.613&    4&   0.100&   0.114&   0.104&   0.109&   0.112\\ 
&   2612.65&   384.790&    8&   0.110&   0.131&   0.106&   0.125&   0.125\\ 
&   2607.86&   667.683&    6&   0.110&   0.118&   0.110&   0.118&   0.124\\ 
&   2600.17&     0.000&   10&   0.230&   0.240&   0.224&   0.240&   0.245\\ 
&   2599.14&   384.790&    8&   0.099&   0.095&   0.099&   0.109&   0.109\\ 
&   2586.65&     0.000&   10&   0.068&   0.055&   0.065&   0.069&   0.071\\ 
&   2414.04&   977.053&    2&   0.190&   0.175&   0.186&   0.177&   0.186\\ 
&   2411.80&   977.053&    2&   0.210&   0.220&   0.208&   0.208&   0.223\\ 
&   2411.25&   862.613&    4&   0.190&   0.203&   0.194&   0.208&   0.223\\ 
&   2407.39&   862.613&    4&   0.140&   0.161&   0.141&   0.147&   0.161\\ 
&   2405.61&   667.683&    6&   0.200&   0.243&   0.196&   0.235&   0.252\\ 
&   2405.16&   862.613&    4&   0.031&   0.031&   0.031&   0.026&   0.028\\ 
&   2399.97&   667.683&    6&   0.120&   0.131&   0.118&   0.118&   0.124\\ 
&   2396.35&   384.790&    8&   0.270&   0.288&   0.261&   0.286&   0.307\\ 
&   2396.14&   667.683&    6&   0.019&   0.023&   0.019&   0.015&   0.016\\ 
&   2389.35&   384.790&    8&   0.089&   0.097&   0.088&   0.083&   0.088\\ 
&   2383.78&   384.790&    8&   0.012&   0.012&     -  &     -  &     -  \\ 
&   2382.76&     0.000&   10&   0.364&   0.343&   0.417&   0.316&   0.347\\ 
&   2381.48&   667.683&    6&   0.036&   0.010&   0.036&   0.034&   0.036\\ 
&   2374.46&     0.000&   10&   0.028&   0.053&   0.039&   0.032&   0.033\\ 
&   2367.59&     0.000&   10&   0.004&   0.004&     -  &     -  &     -  \\ 
&   2365.55&   384.790&    8&   0.051&   0.055&   0.051&   0.050&   0.053\\ 
&   2359.82&   862.613&    4&   0.038&   0.038&     -  &     -  &     -  \\ 
&   2349.02&   667.683&    6&   0.100&   0.077&   0.103&     -  &   0.090\\ 
&   2345.00&   977.053&    2&   0.130&   0.126&   0.138&   0.158&   0.169\\ 
&   2344.21&     0.000&   10&   0.113&   0.126&   0.110&   0.115&   0.126\\ 
&   2338.72&   862.613&    4&   0.087&   0.087&   0.087&   0.093&   0.100\\ 
&   2333.51&   384.790&    8&   0.091&   0.081&   0.097&     -  &   0.077\\ 
&   2328.11&   667.683&    6&   0.032&   0.040&   0.031&   0.036&   0.037\\ 
$a^4F$ &   2360.72&  1873&   10&   0.020&   0.040&   0.028&   0.028&   0.023\\ 
&   2348.83&  1873&   10&   0.034&   0.086&   0.037&   0.043&   0.037\\ 
&   2332.02&  1873&   10&   0.019&   0.004&   0.021&   0.021&   0.018\\ 
&   2369.32&  2838&    6&   0.033&   0.064&   0.033&   0.034&   0.030\\ 
&   2361.02&  2430&    8&   0.037&   0.073&   0.037&     -  &   0.031\\ 
$a^4D$ &   2564.24&  8392&    6&   0.085&   0.098&   0.084&     -  &   0.094\\ 
&   2563.30&  7955&    8&   0.110&   0.143&   0.111&     -  &   0.134 \\
\enddata
\tablenotetext{a}{NIST database, mostly Fuhr, Martin \& Wiese 1988,
J. Phys. Chem. Ref. Data, 17, Suppl. 4, 108}
\tablenotetext{b}{Iron Project, Nahar 1995, A \& A 293, 967}
\tablenotetext{c}{Compilation of experimental data, Giridhar \& Ferro 1995, Rev. Mex. 
Astron. Astrofis., 31, 23}
\tablenotetext{d}{Experimental, Bergeson et al. 1996, ApJ, 464, 1044 }
\tablenotetext{e}{Raassen \& Uylings 1998, A \& A, 340, 300 }

\end{deluxetable}

\begin{deluxetable}{cccc}
\tablewidth{9cm}
\tablecaption{\sc Relative population of \feii\ energy levels \label{levpop}} 
\tablehead{
\colhead{Term}
&\colhead{Energy [cm$^{-1}$]}
&\colhead{$\zeta$(1)}
&\colhead{$\zeta$(2)}
}
\startdata
$a^6D$ &    0. &  1.00 &   1.00 \\
&  384. &  0.04 (0.02)&   0.12 (0.03)\\
&  667. &  0.04 (0.02)&   0.10 (0.03)\\
&  862. &  0.04 (0.03)&  0.10 (0.03)\\
&  977. &  $<$0.1  &  0.10 (0.03)\\
\\[0.01cm]
$a^4F$ & 1873. &  $<$0.3  &  0.50 (0.1)\\
& 2430. &  $<$0.6  &  $<$0.2  \\
& 2838. &  $<$0.6  &  $<$0.2 \\
\\[0.01cm]
$a^4D$ & 7955. &  $<$0.1  &  $<$0.03 \\ 
&  8392. &  $<$0.2  &  $<$0.1  \\
\enddata
\tablecomments{The relative population per unit statistical weight of the
low-lying energy levels of \feii\ that were derived from the relative
strengths of the absorption lines.  $\zeta$ would be equal to 1 if
the level population were in LTE at a temperature higher than the level
energy.  $\zeta$(1) applies to the low velocity line
cluster, $\zeta$(2) to the high velocity
cluster. Note that the errors only reflect the
fitting uncertainty, and do not include systematic errors like
uncertainties in the continuum.
}
\end{deluxetable}

\begin{deluxetable}{ccccccccccc}
\rotate 90
\tabletypesize{\footnotesize}
\tablewidth{20cm}
\tablecaption{\sc Ionization models \label{ionmods}}
\tablehead{
\colhead{Model \#}
&\colhead{$\log U$}
&\colhead{$\log n_H$}
&\colhead{spectrum}\tablenotemark{a}
&\colhead{abundance}\tablenotemark{b}
&\colhead{dust}\tablenotemark{b}
&\colhead{$\log N_H$}\tablenotemark{c}
&\colhead{$\log N_{Mg~I}$}\tablenotemark{d}
&\colhead{$\log N_{Fe~I}$}\tablenotemark{e}
&\colhead{$A_{2500}$}\tablenotemark{f}
&\colhead{Distance (kpc)}\tablenotemark{g}
}
\startdata
1&   -1.0 & 3.90 & AGN & solar  &  no & 22.00 & 12.45 & 10.44 & - & 0.24 \\
2&   -2.0 & 4.00 & AGN & solar & no & 21.01 & 12.89 & 11.23  & - & 0.67\\
3&   -3.0 & 4.10 & AGN & solar & no & 20.25 & 12.75 & 11.66  & - & 1.9 \\
4&   -4.0 & 4.20 & AGN & solar & no & ? & $>$12.47 & $>$12.09  & - & 5.3 \\
5&   -5.0 & 4.50 & AGN & solar & no & ? & $>$12.82 & $>$12.56  & - & 19 \\
6&   -6.0 & 4.90 & AGN & solar & no & ? & $>$12.82 & $>$13.20  & - & 94\\
7&   -2.0 & 4.00 & AGN & 3 $\times$ solar & no & 20.97 & 13.35 & 11.48 	& - & 0.67 \\
8&   -2.0 & 4.00 & power law& solar  & no & 21.03 & 12.74 & 11.21 & - & 0.53\\
9&   -4.0 & 4.20 & power law& solar  & no & 20.00 & 12.34 & 12.00 & - & 4.2\\
10&  -6.0 & 4.90 & power law& solar  & no & ? & $>$12.79 & $>$13.06 & - & 74 \\ 
11&  -1.0 & 4.60 & AGN & Orion & yes  & 21.70 & 12.61 & 11.88 & 4.2& 0.10\\
12&  -2.0 & 5.00 & AGN & Orion & yes  &21.26 & 12.49 & 11.67 & 1.5& 0.21\\
13&  -3.0 & 5.00 & AGN & Orion & yes  &21.06 & 12.86 & 12.30 & 0.96& 0.66\\
14&  -4.0 & 5.00 & AGN & Orion & yes  & ? & $>$12.65 & $>$12.42 & ? & 2.1\\
15&  -3.0 & 5.00 & AGN & Orion Mg $\times$ 3 & yes  &21.06 & 13.34 & 12.30 & 0.96 & 0.66\\
\enddata
\tablenotetext{a}{AGN refers to Matthews \& Ferland, 1987}
\tablenotetext{b}{Abundances and dust properties based on values for
the Orion nebula, Baldwin et al. 1991,Osterbrock et al. 1992}
\tablenotetext{c}{Total hydrogen column density up to $\log N_{Fe~II}=15.5$}
\tablenotetext{d}{Total \mgi\ column density up to $\log N_{Fe~II}=15.5$}
\tablenotetext{e}{Total \fei\ column density up to $\log N_{Fe~II}=15.5$}
\tablenotetext{f}{Expected dust extinction at 2500 \AA\ for the models
with dust}
\tablenotetext{g}{Distance implied by density and ionization parameter
assuming a QSO luminosity of $10^{46} {\rm erg s^{-1}}$}
\end{deluxetable}

\begin{figure}
\plotone{fig1.eps}
\end{figure}

\begin{figure}
\plotone{fig2.eps}
\end{figure}

\begin{figure}
\plotone{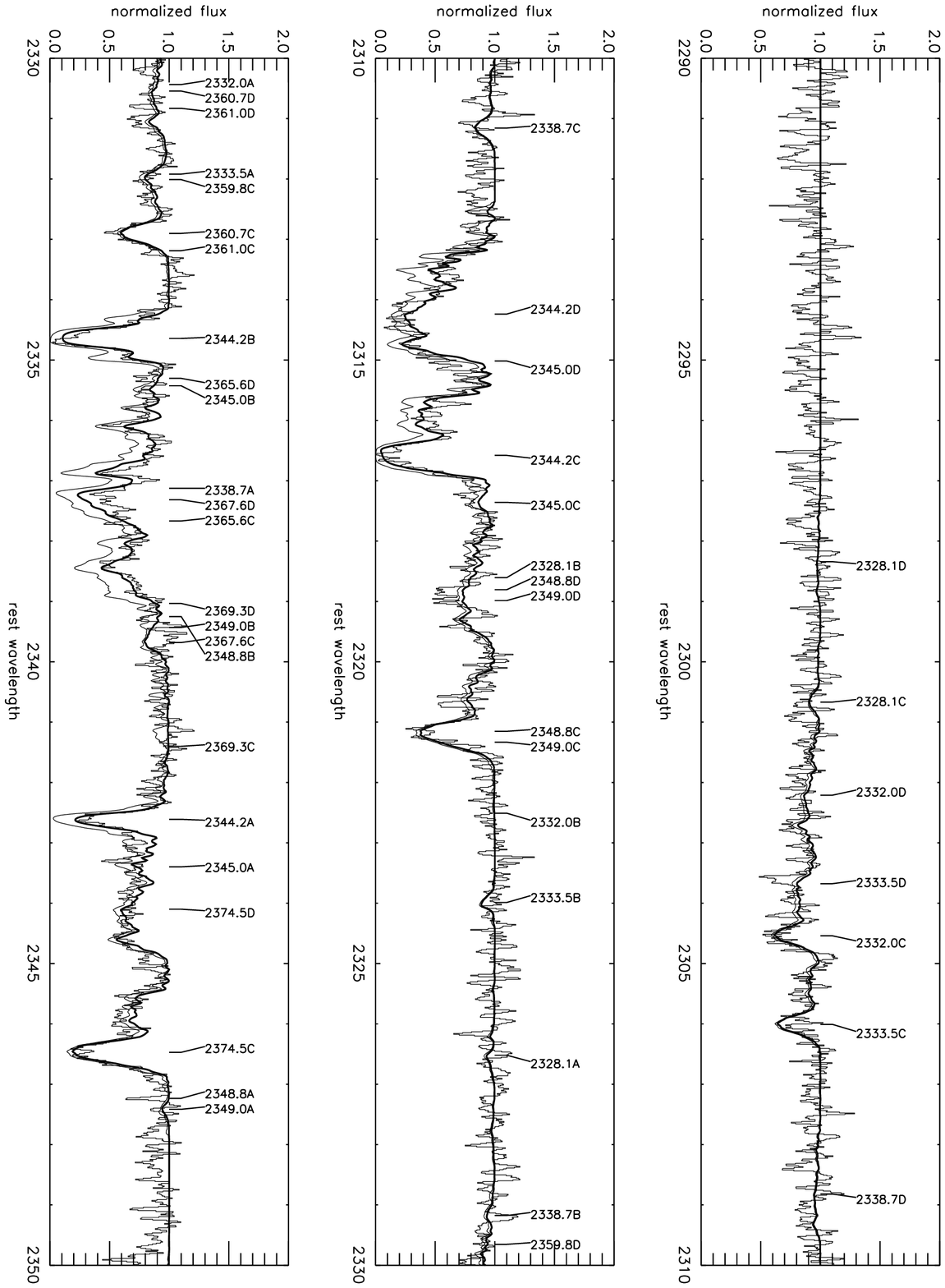}
\end{figure}

\begin{figure}
\plotone{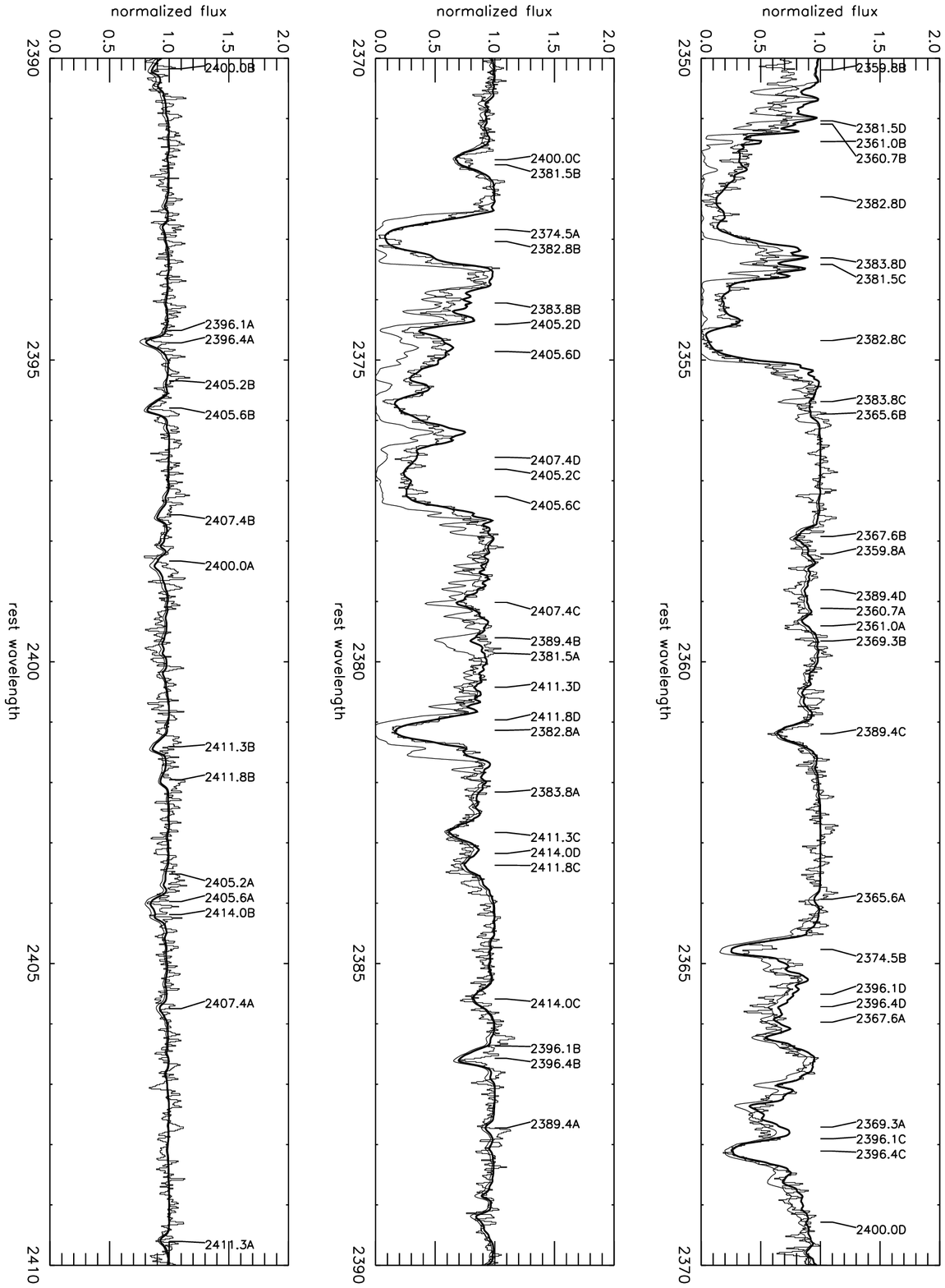}
\end{figure}

\begin{figure}
\plotone{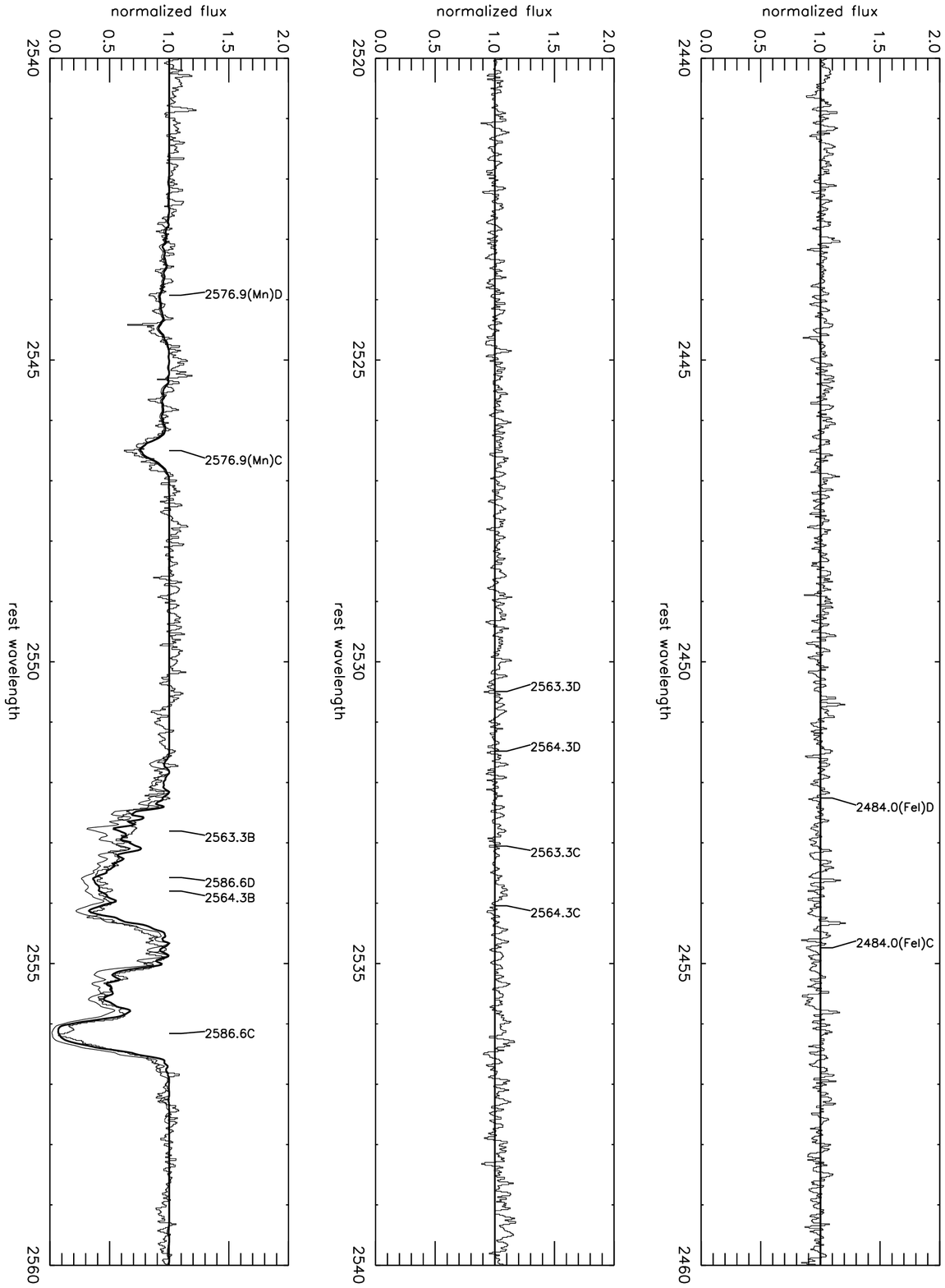}
\end{figure}

\begin{figure}
\plotone{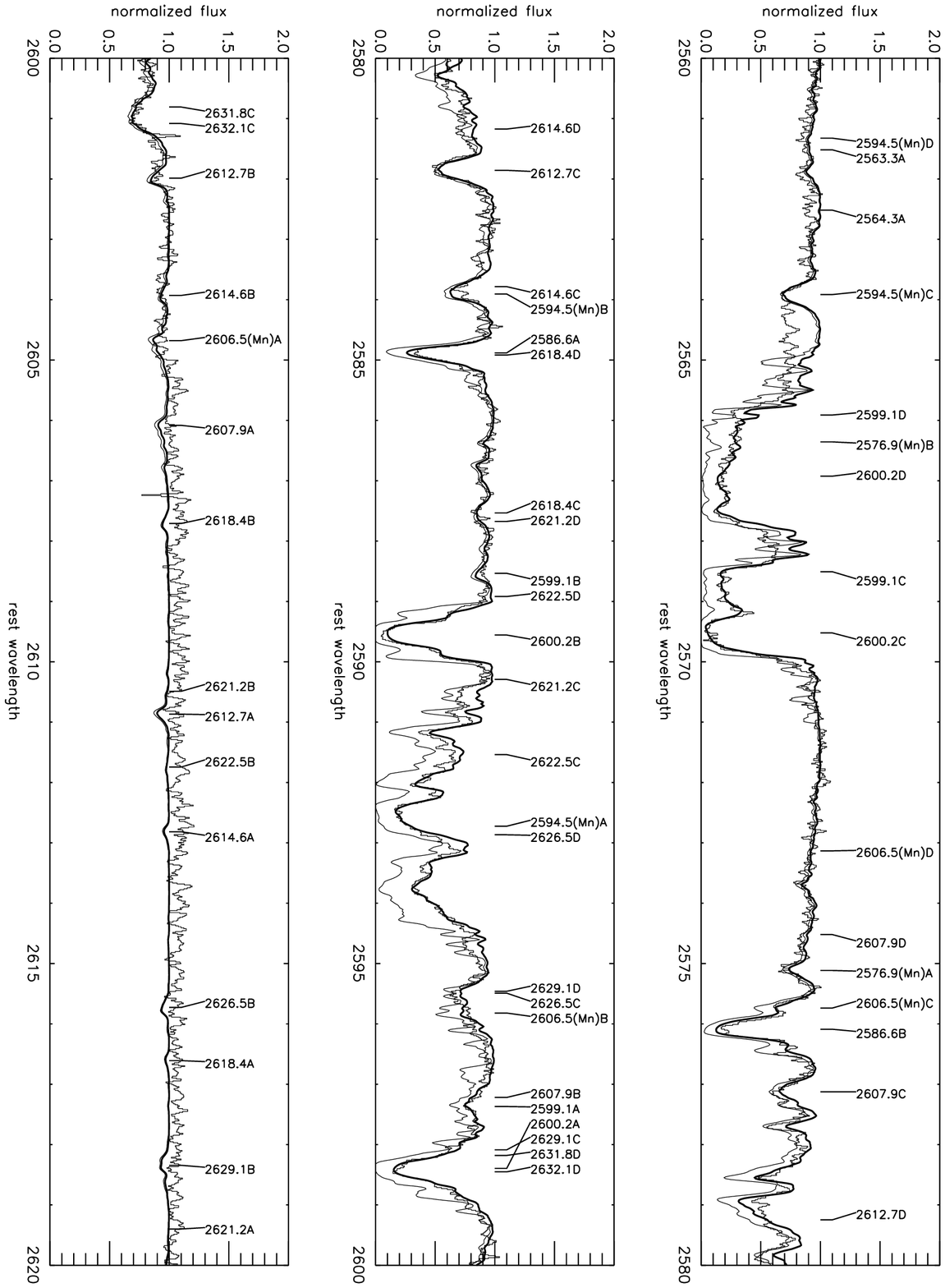}
\end{figure}

\begin{figure}
\plotone{fig4.eps}
\end{figure}

\begin{figure}
\plotone{fig5.eps}
\end{figure}

\begin{figure}
\epsscale{0.9}
\plotone{fig6.eps}
\end{figure}

\begin{figure}
\epsscale{0.9}
\plotone{fig7.eps}
\end{figure}

\begin{figure}
\epsscale{0.9}
\plotone{fig8.eps}
\end{figure}


\newpage
\begin{references}
\reference{} Arav, N., \& Li, Z. Y. 1994, ApJ, 427, 700
\reference{} Arav, N., Korista, K.T., de Kool, M., Junkkarinen, V.T. \&
Begelman, M.C. 1999a, \apj, 516, 27
\reference{} Arav, N., Becker, R.H., Laurent-Muehleisen, S.A., 
Gregg, M.D., White, R.L. \& de Kool, M. 1999b, \apj, 524,566.
\reference{} Baldwin, J., Ferland, G.J., Martin, P.G., Corbin, M., Cota, S., Peterson, B.M., \& Slettebak, A., 1991, ApJ 374, 580
\reference{} Barlow, T.A. 2000, AJ, in preparation
\reference{} Barlow, T.A. \& Sargent, W.L.W. 1997, AJ, 113, 136 
\reference{} Begelman, M.C., de Kool, M. \& Sikora, M. 1991, \apj,
382,416
\reference{} Bergeson et al. 1996, ApJ, 464, 1044
\reference{} Cardelli, J.A. \& Savage, B.D. 1995, \apj, 452, 275
\reference{} Collin-Souffrin, S., Joly, M., Pequignot, D. \& Dumont, S. 1986, 
A \& A, 166, 27
\reference{} Ferland, G.J., 1996, Hazy, a Brief Introduction to Cloudy, 
University of Kentucky Department of Physics and Astronomy Internal
Report.
\reference{} Foltz, C.B. et al. 1986, \apj, 307,504
\reference{} Francis, P.J., Hewett, P.C., Foltz, C.B., Chafee, F.H., 
Weymann, R.J. \& Morris, S.L. 1991, \apj, 373, 465 
\reference{} Fuhr, R.J., Martin,G.A. \& Wiese,W.L. 1988, J. Phys. Chem. Ref. Data, 17, Suppl. 4, 108
\reference{} Ganguly, R., Eracleous, M., Charlton, J.C. \& Churchill,
C.W. 1999, \apj, 117, 2594
\reference{} Giridhar, S. \& Ferro, A.F. 1995, Rev. Mex. 
Astron. Astrofis., 31, 23
\reference{} Hamann, F., Korista, T. K., \& Morris, S. L. 1993 \apj, 415, 541
\reference{} Hamann, F., Barlow, T.A., Beaver, E.A., Burbidge, E.M., 
Cohen, R.D., Junkkarinen, V.T. \& Lyons., R. 1995, \apj, 443, 606 
\reference{} Hamann, F., Barlow, T.A., \& Junkkarinen, V.T. 1997, \apj, 478,
87
\reference{} Hamann, F.W., Netzer, H. \& Shields, J.C. 2000, \apj,
536, 101
\reference{} Laor, A., Fiore, F., Elvis,M., Wilkes, B.J. \& McDowell,
J.C. 1997, \apj, 477, 93
\reference{} Mathews,W.G. \& Ferland, G. 1987, \apj, 323, 456
\reference{} Murray, N., Chiang, J., Grossman, S.A., \& Voit,
G.M. 1995, \apj, 451, 498
\reference{} Nahar, S.N. 1995, A \& A 293, 967
\reference{} Osterbrock, D. E., Tran, H.D., \& Veilleux, S., 1992, ApJ 389, 305
\reference{} Petitjean, P. \& Srianand, R. 1999, A \& A, 345, 73
\reference{} Raassen, A.J.J. \& Uylings, P.H.M. 1998, A \& A, 340, 300
\reference{} Srianand,R. \& Petitjean, P. 2000, A \& A, 357, 414
\reference{} Turnshek, D. A. 1988, in Space Telescope Sci. Inst. Symp. 2, QSO
Absorption Lines: Probing the Universe, ed. S. C. Blades, D. A.
Turnshek, \& C. A. Norman (Cambridge: Cambridge Univ. Press), 17
\reference{} Verner, E.M., Verner, D.A., Korista, K.T., Ferguson,
J.W., Hamann, F., \& Ferland, G.J. 1999, \apjs, 120, 101 
\reference{} Vogt et al. 1994 
\reference{} Voit, G.M., Weymann, R.J. \& Korista, K.T. 1993, \apj, 413, 95
\reference{} Wampler, E.J., Chugai, N.N. \& Petitjean, P. 1995, 
\apj, 443, 586 (WCP)
\reference{} Weymann, R.J., Williams, R.E., Peterson, B.M. \&
Turnshek, D.A. 1979, \apj, 234, 33
\reference{} Weymann, R. J., Turnshek, D. A., \& Christiansen, W. A. 1985, in
Astrophysics of Active Galaxies and Quasi-stellar Objects, ed. J. Miller
(Oxford: Oxford Univ. Press) 333 (WTC)
\reference{} Weymann, R. J., Morris, S. L., Foltz, C.B. \& Hewett, P.C. 1991,
\apj, 373, 23
\reference{} Weymann R.J. 1997, in ASP Conf. Ser. 128, Masss Ejection
from AGN, ed. N. Arav, I. Shlosman, \& R.J. Weymann (San Francisco:ASP),3
\reference{} White, R.L. et al. 2000, ApJS, 126, 133
\reference{} Wills, B.J., Netzer, H, \& Wills, D. 1985, \apj, 288, 94
\reference{} Zhang, H.L. \& Pradhan, A.K. 1995, A \& A, 293, 953
H. Nussbaumer \& P. J. Storey, 1980, A \& A , 89, 308
Quinet P., Le Dourneuf M. \& Zeippen C.J., 1996, A \& A Suppl. 120, 361 

\end{references}
\end{document}